\documentclass[aps,onecolumn,prd,preprintnumbers,superscriptaddress,nobibnotes,floatfix,longbibliography,nofootinbib]{revtex4-2}

\pdfoutput=1
\usepackage{amsmath,amsfonts,amssymb,mathrsfs,graphicx,color,longtable}
\usepackage{hyperref}
\usepackage{bm}
\usepackage{array}
\usepackage{color}
\usepackage{enumitem}
\usepackage[explicit]{titlesec}
\usepackage{ulem}
\usepackage[utf8]{inputenc}
\usepackage{float}  
\usepackage{multirow,bigdelim}
\usepackage{blindtext}
\usepackage{rotating}
\usepackage[ddmmyy,24hr]{datetime}

\usepackage[TABBOTCAP,tight]{subfigure}
\newcolumntype{C}[1]{>{\centering\let\newline\\\arraybackslash\hspace{4pt}}m{#1}}

\graphicspath{{}}

\begin{document}

\title{Inspection of the detection cross section dependence of the Gallium Anomaly}

\author{C. Giunti}
\email{carlo.giunti@to.infn.it}
\affiliation{Istituto Nazionale di Fisica Nucleare (INFN), Sezione di Torino, Via P. Giuria 1, I--10125 Torino, Italy}

\author{Y.F. Li}
\email{liyufeng@ihep.ac.cn}
\affiliation{Institute of High Energy Physics, Chinese Academy of Sciences, Beijing 100049, China}
\affiliation{School of Physical Sciences, University of Chinese Academy of Sciences, Beijing 100049, China}

\author{C.A. Ternes}
\email{ternes@to.infn.it}
\affiliation{Istituto Nazionale di Fisica Nucleare (INFN), Sezione di Torino, Via P. Giuria 1, I--10125 Torino, Italy}
\affiliation{Dipartimento di Fisica, Universit\`a di Torino, via P. Giuria 1, I--10125 Torino, Italy}

\author{Z. Xin}
\email{xinzhao@ihep.ac.cn}
\affiliation{Institute of High Energy Physics,
Chinese Academy of Sciences, Beijing 100049, China}
\affiliation{School of Physical Sciences, University of Chinese Academy of Sciences, Beijing 100049, China}

%\date{\dayofweekname{\day}{\month}{\year} \ddmmyydate\today, \currenttime}

\begin{abstract}
We discuss in detail the dependence of the Gallium Anomaly on the detection cross section.
We provide updated values of the size of the Gallium Anomaly and find that its significance is larger than about $5\sigma$ for all the detection cross section models.
We discuss the dependence of the Gallium Anomaly on the assumed value of the half life of ${}^{71}\text{Ge}$,
which determines the cross sections of the transitions
from the ground state of
${}^{71}\text{Ga}$ to the ground state of ${}^{71}\text{Ge}$.
We show that a value of the ${}^{71}\text{Ge}$ half life
which is larger than the standard one can reduce or even solve the Gallium Anomaly.
Considering the short-baseline neutrino oscillation interpretation of the Gallium Anomaly,
we show that a value of the ${}^{71}\text{Ge}$ half life
which is larger than the standard one can reduce the tension with the results of
other experiments.
Since the standard value of the ${}^{71}\text{Ge}$ half life was measured in 1985,
we advocate the importance of new measurements
with modern technique and apparatus for a better assessment of the Gallium Anomaly.
\end{abstract}

\maketitle
%\tableofcontents

%%%%%%%%%%%%%%%%%%%%%%%%%%%%%%%%%%%%%%%%%%%%%%%%%%%%%%%%

\section{Introduction}
\label{sec:intro}

The Gallium Anomaly is one of the current major puzzles in neutrino physics.
It consists of a relatively large deficit of the rate of events observed in
Gallium source experiments
with respect to the expectation.
It was initially discovered~\cite{Abdurashitov:2005tb,Laveder:2007zz,Giunti:2006bj}
in analyses of the data of the
GALLEX~\cite{GALLEX:1994rym,GALLEX:1997lja,Kaether:2010ag}
and
SAGE~\cite{Abdurashitov:1996dp,SAGE:1998fvr,Abdurashitov:2005tb,SAGE:2009eeu}
radioactive source experiments
which have been performed as tests of the solar neutrino detection done in these experiments
through the process
\begin{equation}
\nu_e + {}^{71}\text{Ga} \to {}^{71}\text{Ge} + e^-
.
\label{eq:detproc}
\end{equation}
The GALLEX collaboration performed two experiments
with a ${}^{51}\text{Cr}$ radioactive source which produced electron neutrinos
through the electron capture (EC) process
$e^- + {}^{51}\text{Cr} \to {}^{51}\text{V} + \nu_e$.
There are four neutrino energy lines,
which are shown in Fig.~2 of Ref.~\cite{Bahcall:1997eg},
and the average neutrino energy is 0.72~MeV.
The SAGE collaboration performed an experiment
with a ${}^{51}\text{Cr}$ radioactive source
and another experiment with a 
${}^{37}\text{Ar}$ radioactive source
which emitted electron neutrinos through the EC process
$e^- + {}^{37}\text{Ar} \to {}^{37}\text{Cl} + \nu_e$.
In this case, there are two neutrino energy lines,
which are shown in Fig.~2 of Ref.~\cite{Abdurashitov:2005tb},
and the average neutrino energy is 0.81~MeV.
The average ratio of measured and predicted ${}^{71}\text{Ge}$ production rates
was found to be
$\overline{R}_{\text{GALLEX+SAGE}}^{\text{Bahcall}} = 0.88 \pm 0.05$~\cite{Abdurashitov:2005tb},
using the traditional Bahcall detection cross section~\cite{Bahcall:1997eg}.
The deviation of $\overline{R}_{\text{GALLEX+SAGE}}$
from unity by about $2.4\sigma$ was called the ``Gallium Anomaly''~\cite{Giunti:2006bj}
and was studied as a possible indication of active-sterile neutrino mixing
(see the reviews in Refs.~\cite{Gariazzo:2015rra,Gonzalez-Garcia:2015qrr,Giunti:2019aiy,Diaz:2019fwt,Boser:2019rta,Dasgupta:2021ies}).

A recent check of the Gallium Anomaly
has been performed in the BEST experiment~\cite{Barinov:2021asz,Barinov:2022wfh}
by using a ${}^{51}\text{Cr}$ radioactive source
placed at the center of a detector with two nested
${}^{71}\text{Ga}$ volumes
corresponding to average neutrino path lengths of about
$0.5~\text{m}$ and $1.1~\text{m}$.
Assuming the traditional Bahcall detection cross section~\cite{Bahcall:1997eg},
the ratios of measured and predicted ${}^{71}\text{Ge}$ production rates
in the inner and outer BEST volumes are
$\overline{R}_{\text{BEST-in}}^{\text{Bahcall}} = 0.79 \pm 0.05$
and
$\overline{R}_{\text{BEST-out}}^{\text{Bahcall}} = 0.77 \pm 0.05$~\cite{Barinov:2021asz,Barinov:2022wfh} respectively,
which lead to a total average ratio
$\overline{R}_{\text{GALLEX+SAGE+BEST}}^{\text{Bahcall}} = 0.80 \pm 0.05$~\cite{Barinov:2021asz,Barinov:2022wfh}
or
$\overline{R}_{\text{GALLEX+SAGE+BEST}}^{\text{Bahcall}} = 0.80 \pm 0.04$~\cite{Giunti:2022btk}.
The slight difference of these average values is due to small differences in the treatment of the correlated uncertainties.
In any case,
it is clear that the BEST measurements confirmed the Gallium Anomaly
and increased its statistical significance to a level of about
$4$-$5\sigma$
(assuming the traditional Bahcall detection cross section)~\cite{Giunti:2022btk}.

The Gallium Anomaly can be due to the disappearance of
the electron neutrinos during their propagation from the source to the
detection point.
Such disappearance can naturally be caused by neutrino oscillations,
which are known to exist and are due to neutrino masses
(see, e.g., the Particle Data Group review in Ref.~\cite{ParticleDataGroup:2022pth}).
However,
in the GALLEX, SAGE, and BEST radioactive source experiments
the average neutrino path length $L$ was rather short
(about $1.9~\text{m}$ for GALLEX, $0.6~\text{m}$ for SAGE and less than $1.1~\text{m}$ for BEST).
Taking into account the average neutrino energy $E$ of about $0.7-0.8$~MeV,
the oscillations must be due to a squared-mass difference
$\Delta{m}^2 \gtrsim E/L \sim 1~\text{eV}^2$,
which is much larger than the well established squared-mass differences
of solar and atmospheric neutrino oscillations
in the standard three-neutrino mixing framework~\cite{deSalas:2020pgw,Capozzi:2021fjo,Esteban:2020cvm},
where the three known active flavor neutrinos
$\nu_{e}$,
$\nu_{\mu}$, and
$\nu_{\tau}$
are unitary linear combinations of three massive neutrinos
$\nu_{1}$,
$\nu_{2}$, and
$\nu_{3}$
with respective masses
$m_{1}$,
$m_{2}$, and
$m_{3}$.
Therefore,
a neutrino oscillation explanation of the Gallium Anomaly
and other short-baseline (SBL) neutrino anomalies~\cite{Gariazzo:2015rra,Gonzalez-Garcia:2015qrr,Giunti:2019aiy,Diaz:2019fwt,Boser:2019rta,Dasgupta:2021ies}
requires the introduction of at least an additional massive neutrino
$\nu_{4}$
with mass $m_{4} \gtrsim 1 \, \text{eV}$,
which generates the new short-baseline
squared-mass difference
$\Delta{m}^2_{41} \equiv m_{4}^2 - m_{1}^2 \gtrsim 1~\text{eV}^2$,
assuming $m_{1}, m_{2}, m_{3} \ll 1 \, \text{eV}$
as indicated by $\beta$ decay, neutrinoless double-$\beta$ decay and cosmological bounds~\cite{Gariazzo:2015rra,Gonzalez-Garcia:2015qrr,Giunti:2019aiy,Diaz:2019fwt,Boser:2019rta,Dasgupta:2021ies}.
In the flavor basis,
the new neutrino state corresponds to a sterile neutrino,
because the LEP measurements of the invisible width of the $Z$ boson
have shown that there are only three active neutrinos~\cite{ParticleDataGroup:2022pth}.

In the neutrino oscillation framework there is a rather strong tension between the Gallium data and other neutrino data~\cite{Berryman:2021yan,Giunti:2022btk}. In order to explain the Gallium anomaly with neutrino oscillations a rather large mixing angle $\sin^22\vartheta_{ee}\gtrsim 0.2$ is required, independently of the assumed cross section model. As we have shown in Ref.~\cite{Giunti:2022btk}, large mixing angles are disfavored from the analyses of reactor rates~\cite{Giunti:2021kab}, reactor spectral ratios and solar neutrino data (and also to some extent from Tritium $\beta$-decay data)~\cite{Giunti:2022btk}. Interestingly, if all data are analyzed in a combined way (excluding Gallium data) there is a preference for short baseline oscillations, which, depending on the exact data set and reactor flux model considered, is at the 2.7--3.3$\sigma$ level. However, these data prefer small mixing angles, with the best fit found at $\sin^22\vartheta_{ee}\sim 0.02$ and the upper limit $\sin^22\vartheta_{ee}\lesssim 0.1$ at 3$\sigma$. It is clear that the Gallium data cannot be accommodated in a combined way with the other neutrino data discussed in Ref.~\cite{Giunti:2022btk}\footnote{It should be noted that also the Neutrino-4 collaboration found some evidence of neutrino oscillations with large mixing angle~\cite{Serebrov:2020kmd}. These results are not in tension with the ones from Gallium experiments~\cite{Barinov:2021mjj}. However, the data analysis of the Neutrino-4 collaboration has been questioned in Refs.~\cite{Danilov:2020rax,Giunti:2021iti}.}.

Let us also emphasize that
after the results of the BEST experiment
the Gallium Anomaly is still based on the absolute
comparison of the observed and predicted rates,
because the almost equal values of the ratios of measured and predicted ${}^{71}\text{Ge}$ production rates
in the inner and outer BEST volumes
do not provide any indication of a variation of the $\nu_{e}$ flux
with distance, which would be a model-independent evidence of
neutrino oscillations.

Another possible explanation of the Gallium Anomaly
is an overestimation of the detection cross section.
However, this overestimation of the detection cross section
cannot be due to an overestimation of the
calculated cross sections of the transitions
from the ground state of
${}^{71}\text{Ga}$
to the accessible excited energy levels of ${}^{71}\text{Ge}$,
because the Gallium Anomaly persists even assuming the
absence of these transitions~\cite{Berryman:2021yan,Giunti:2022btk}.
The contribution from the excited states to the total cross section can only be at the few percent level, and even when it is neglected the significance of the anomaly is at the 5$\sigma$ level.

In this paper we inspect the dependence of the Gallium Anomaly
on the cross sections of the transitions
from the ground state of
${}^{71}\text{Ga}$ to the ground state of ${}^{71}\text{Ge}$,
which has been so far assumed to be
known with very small uncertainty
because it is derived from the measured half life of ${}^{71}\text{Ge}$.
However, several measurements of this half life have been performed in the past and the results differ by about 17\% (see Subsection~\ref{sec:lifetime} and Eqs.~\eqref{eq:BGZZ}--\eqref{eq:HR}).

Section~\ref{sec:detection} is dedicated to the effects of the detection cross section.
In the beginning we briefly review the detection cross section models.
We also introduce a new $\chi^2$ function which allows to analyze the Gallium data
taking into account the lower bound on the cross section in each model
given by the transition from the ground state of
${}^{71}\text{Ga}$ to the ground state of ${}^{71}\text{Ge}$.
In Sections~\ref{sec:lifetime} and~\ref{sec:dependence} we discuss the dependence of the Gallium anomaly on the half life of ${}^{71}\text{Ge}$.
In Section~\ref{sec:oscillations}
we show that a value of the half life of ${}^{71}\text{Ge}$
which is larger than the standard one
can reduce the tension between the Gallium data and other data
in the neutrino oscillation framework.
Finally, in Section~\ref{sec:conclusions} we draw our conclusions.

\begin{table*}
\setlength{\tabcolsep}{5pt}
\centering
\begin{tabular}{lcccc}
Model
&
Method
&
$\dfrac{\text{BGT}_{5/2^{-}}}{\text{BGT}_{\text{gs}}^{\text{HR}}}$
&
$\dfrac{\text{BGT}_{3/2^{-}}}{\text{BGT}_{\text{gs}}^{\text{HR}}}$
&
$\dfrac{\text{BGT}_{5/2^{+}}}{\text{BGT}_{\text{gs}}^{\text{HR}}}$
\\
\hline
Bahcall (1997)~\cite{Bahcall:1997eg}
&
$ {}^{71}\text{Ga} (p,n) {}^{71}\text{Ge} $
&
$0.028 \pm 0.009$
&
$0.146 \pm 0.049$
&
$-$
\\
Haxton (1998)~\cite{Haxton:1998uc}
&
{Shell Model}
&
$0.19 \pm 0.18$
&
$-$
&
$-$
\\
Frekers et al. (2015)~\cite{Frekers:2015wga}
&
$ {}^{71}\text{Ga} ({}^{3}\text{He},{}^{3}\text{H}) {}^{71}\text{Ge} $
&
$0.040 \pm 0.031$
&
$0.207 \pm 0.016$
&
$-$
\\
Kostensalo et al. (2019)~\cite{Kostensalo:2019vmv}
&
{Shell Model}
&
$0.033 \pm 0.017$
&
$0.016 \pm 0.008$
&
$( 4.46 \pm 2.24 ) \times 10^{-6} $
\\
Semenov (2020)~\cite{Semenov:2020xea}
&
$ {}^{71}\text{Ga} ({}^{3}\text{He},{}^{3}\text{H}) {}^{71}\text{Ge} $
&
$0.039 \pm 0.030$
&
$0.204 \pm 0.016$
&
$-$
\\
\end{tabular}
\caption{\label{tab:BGT}
Values of the BGT ratios in Eq.~\eqref{eq:cs} of the transitions to the excited states of ${}^{71}\text{Ge}$
in the different cross section models.
The method used for the derivation of the BGT ratios
in each cross section model is indicated in the second column.
}
\end{table*}

\section{Detection Cross Section}
\label{sec:detection}

The interpretation of the data of the Gallium source experiments
depends on the prediction,
which is based on the evaluation of the neutrino flux from the activity of the source
and the detection cross section.
The total cross sections of the detection process~\eqref{eq:detproc}
for electron neutrinos produced by the electron capture decays of
${}^{51}\text{Cr}$ and ${}^{37}\text{Ar}$
are given by~\cite{Bahcall:1997eg,Kostensalo:2019vmv}
\begin{equation}
\sigma_{\text{tot}}
=
\sigma_{\text{gs}}
\left(
1
+
\xi_{5/2^{-}}
\dfrac{\text{BGT}_{5/2^{-}}}{\text{BGT}_{\text{gs}}}
+
\xi_{3/2^{-}}
\dfrac{\text{BGT}_{3/2^{-}}}{\text{BGT}_{\text{gs}}}
+
\xi_{5/2^{+}}
\dfrac{\text{BGT}_{5/2^{+}}}{\text{BGT}_{\text{gs}}}
\right)
,
\label{eq:cs}
\end{equation}
where $\sigma_{\text{gs}}$
is the cross section of the transition from the ground state of
${}^{71}\text{Ga}$
with spin-parity $3/2^{-}$
to the ground state of
${}^{71}\text{Ge}$
with spin-parity $1/2^{-}$.
Because of the spin change,
it is a pure Gamow-Teller transition,
which is determined by the Gamow-Teller strength
$\text{BGT}_{\text{gs}}$.
The other contributions in Eq.~\eqref{eq:cs}
are the cross sections of the transitions
from the ground state of
${}^{71}\text{Ga}$
to the accessible energy levels of ${}^{71}\text{Ge}$
with spin-parities
$5/2^{-}$ (175 keV),
$3/2^{-}$ (500 keV), and
$5/2^{+}$ (525 keV).
The last one is possible only for the higher energy ${}^{37}\text{Ar}$ neutrinos,
as substantiated by the phase-space coefficients 
\begin{align}
\null & \null
\xi_{5/2^{-}}({}^{51}\text{Cr}) =0.663,
\null & \null
\null & \null
\xi_{3/2^{-}}({}^{51}\text{Cr}) =0.221,
\null & \null
\null & \null
\xi_{5/2^{+}}({}^{51}\text{Cr}) = 0,
\label{eq:xiCr}
\\
\null & \null
\xi_{5/2^{-}}({}^{37}\text{Ar}) =0.691,
\null & \null
\null & \null
\xi_{3/2^{-}}({}^{37}\text{Ar}) =0.262,
\null & \null
\null & \null
\xi_{5/2^{+}}({}^{37}\text{Ar}) =0.200.
\label{eq:xiAr}
\end{align}
As explained in the following,
the ground state transition $\sigma_{\text{gs}}$ is usually assumed to be known with very small uncertainty.
On the other hand,
the Gamow-Teller strengths
$\text{BGT}_{5/2^{-}}$,
$\text{BGT}_{3/2^{-}}$, and
$\text{BGT}_{5/2^{+}}$
have a relevant uncertainty,
because they need to be calculated from measurements of
${}^{71}\text{Ga} (p,n) {}^{71}\text{Ge}$~\cite{Krofcheck:1985fg,Krofcheck-PhD-1987,Bahcall:1997eg}
or
${}^{71}\text{Ga} ({}^{3}\text{He},{}^{3}\text{H}) {}^{71}\text{Ge}$~\cite{Frekers:2011zz}
reactions,
or with theoretical models~\cite{Haxton:1998uc,Kostensalo:2019vmv}.

In the standard approach~\cite{Bahcall:1997eg},
the ground state transition $\sigma_{\text{gs}}$
is calculated from the half life of ${}^{71}\text{Ge}$
through the relation~\cite{Semenov:2020xea}
\begin{equation}
\sigma_{\text{gs}}
=
\dfrac{G_{\text{F}}^2 \cos^2\vartheta_{\text{C}}}{\pi}
g_{A}^2
\text{BGT}_{\text{gs}}
\left\langle p_{e} E_{e} F(Z_{\text{Ge}}, E_{e}) \right\rangle
=
\dfrac{\pi^2 \ln2}{m_{e}^5 ft_{1/2}({}^{71}\text{Ge})}
\left\langle p_{e} E_{e} F(Z_{\text{Ge}}, E_{e}) \right\rangle
,
\label{eq:sigmags}
\end{equation}
where
$G_{\text{F}}$
is the Fermi constant,
$\vartheta_{\text{C}}$
is the Cabibbo angle,
$g_{A}$
is the axial coupling constant,
$m_{e}$,
$p_{e}$, and
$E_{e}$
are the electron mass, momentum, and energy respectively,
$F(Z_{\text{Ge}}, E_{e})$
is the Fermi function for the atomic number
$Z_{\text{Ge}} = 32$ and $ft_{1/2}({}^{71}\text{Ge})$ is the $ft$-value for Germanium which depends on the half life
(see, e.g., Ref.~\cite{Bahcall:1978fa}).
The average
$\left\langle p_{e} E_{e} F(Z_{\text{Ge}}, E_{e}) \right\rangle$
is done with respect to the energy lines of the neutrinos emitted in
the electron capture decays of
${}^{51}\text{Cr}$ and ${}^{37}\text{Ar}$,
which determine the electron energy through the relation
$E_{e} = E_{\nu} - \Delta{M}$,
where $E_{\nu}$ is the neutrino energy
and
$
\Delta{M}
=
M_{{}^{71}\text{Ge}} - M_{{}^{71}\text{Ga}}
=
232.443 \pm 0.093 \, \text{keV}
$~\cite{Alanssari:2016itw}.

There are different measurements of the half life of ${}^{71}\text{Ge}$ in the literature:
\begin{align}
\null & \null
T_{1/2}^{\text{BGZZ}}({}^{71}\text{Ge}) = 12.5 \pm 0.1 \, \text{d}
\quad
\text{(Bisi, Germagnoli, Zappa, and Zimmer, 1955)~\protect\cite{Bisi:NC1955}}
,
\label{eq:BGZZ}
\\
\null & \null
T_{1/2}^{\text{R}}({}^{71}\text{Ge}) = 10.5 \pm 0.4 \, \text{d}
\quad
\text{(Rudstam, 1956)~\protect\cite{NSR1956RU45}}
,
\label{eq:R}
\\
\null & \null
T_{1/2}^{\text{GRPF}}({}^{71}\text{Ge}) = 11.15 \pm 0.15 \, \text{d}
\quad
\text{(Genz, Renier, Pengra, and Fink, 1971)~\protect\cite{Genz:1971kv}}
,
\label{eq:GRPF}
\\
\null & \null
T_{1/2}^{\text{HR}}({}^{71}\text{Ge}) = 11.43 \pm 0.03 \, \text{d}
\quad
\text{(Hampel and Remsberg, 1985)~\protect\cite{Hampel:1985zz}}
.
\label{eq:HR}
\end{align}
Note that, although these measurements have different levels of accuracy, they differ by about 17\%, considering the range from 10.5 d to 12.5 d.
Therefore, a more reliable measurement of the ${}^{71}\text{Ge}$ half life is rather demanding and motivates us to consider the dependence of the Gallium Anomaly on the ${}^{71}\text{Ge}$ half life. In previous studies of the Gallium Anomaly, the ${}^{71}\text{Ge}$ half life $T_{1/2}^{\text{HR}}({}^{71}\text{Ge})$ measured by Hampel and Remsberg (HR) was used as the nominal value.
Since this assumption is crucial for the current existence of the Gallium Anomaly,
in the following Sections we discuss the implications for the Gallium Anomaly
of the different values of the ${}^{71}\text{Ge}$ half life.

The existing calculations of the BGT ratios in Eq.~\eqref{eq:cs}
are listed in Table~\ref{tab:BGT}.
We refer to each of them as a ``cross section model''.
Table~\ref{tab:BGT} shows that the transition to the $5/2^{+}$ (525 keV)
energy level of ${}^{71}\text{Ge}$
has been neglected in all the cross section models,
except that of Kostensalo et al.,
where the corresponding BGT value was found to be very small.
Hence, it is plausible that this transition is negligible,
but for completeness we will continue to consider it in the following.

In addition to the cross section models in Table~\ref{tab:BGT},
we consider a ``Ground State'' model~\cite{Berryman:2021yan,Giunti:2022btk},
in which it is assumed that the transitions to the excited states of ${}^{71}\text{Ge}$ are negligible.
This is justified by the differences of the BGT values of the different
cross section models and their uncertainties.
It is an extreme possibility which represents
the lowest value that the detection cross section can have
under the only assumption of the reliability of the transitions
from the ground state of
${}^{71}\text{Ga}$ to the ground state of ${}^{71}\text{Ge}$
based on the measurement~\eqref{eq:HR} of the half life of ${}^{71}\text{Ge}$.

\begin{table*}
\setlength{\tabcolsep}{5pt}
\centering
\begin{tabular}{lcccccccc}
Model
&
$R_{\text{GALLEX-1}}^{\text{HR}}$
&
$R_{\text{GALLEX-2}}^{\text{HR}}$
&
$R_{\text{SAGE-Cr}}^{\text{HR}}$
&
$R_{\text{SAGE-Ar}}^{\text{HR}}$
&
$R_{\text{BEST-in}}^{\text{HR}}$
&
$R_{\text{BEST-out}}^{\text{HR}}$
&
$\overline{R}^{\text{HR}}$
&
GA
\\
\hline
Ground State~\cite{Semenov:2020xea}
&
$1.00 \pm 0.12$
&
$0.85 \pm 0.12$
&
$1.00 \pm 0.13$
&
$0.83 \pm 0.10$
&
$0.83 \pm 0.05$
&
$0.80 \pm 0.05$
&
$0.845^{+0.031}_{-0.031}$
&
$5.0$
\\
%Bahcall (1997)~\cite{Bahcall:1997eg}
Bahcall~\cite{Bahcall:1997eg}
&
$0.95 \pm 0.11$
&
$0.81 \pm 0.11$
&
$0.95 \pm 0.12$
&
$0.79 \pm 0.09$
&
$0.79 \pm 0.05$
&
$0.77 \pm 0.05$
&
$0.804^{+0.037}_{-0.036}$
&
$5.2$
\\
%Haxton (1998)~\cite{Haxton:1998uc}
Haxton~\cite{Haxton:1998uc}
&
$0.86 \pm 0.10$
&
$0.74 \pm 0.10$
&
$0.86 \pm 0.11$
&
$0.72 \pm 0.08$
&
$0.72 \pm 0.05$
&
$0.70 \pm 0.05$
&
$0.731^{+0.088}_{-0.072}$
&
$5.1$
\\
%Frekers et al. (2015)~\cite{Frekers:2015wga}
Frekers et al.~\cite{Frekers:2015wga}
&
$0.93 \pm 0.11$
&
$0.79 \pm 0.11$
&
$0.93 \pm 0.12$
&
$0.77 \pm 0.09$
&
$0.78 \pm 0.05$
&
$0.75 \pm 0.05$
&
$0.789^{+0.033}_{-0.032}$
&
$6.1$
\\
%Kostensalo et al. (2019)~\cite{Kostensalo:2019vmv}
Kostensalo et al.~\cite{Kostensalo:2019vmv}
&
$0.97 \pm 0.11$
&
$0.83 \pm 0.11$
&
$0.97 \pm 0.12$
&
$0.81 \pm 0.09$
&
$0.81 \pm 0.05$
&
$0.78 \pm 0.05$
&
$0.825^{+0.031}_{-0.031}$
&
$5.5$
\\
%Semenov (2020)~\cite{Semenov:2020xea}
Semenov~\cite{Semenov:2020xea}
&
$0.93 \pm 0.11$
&
$0.79 \pm 0.11$
&
$0.93 \pm 0.12$
&
$0.77 \pm 0.09$
&
$0.77 \pm 0.05$
&
$0.75 \pm 0.05$
&
$0.787^{+0.033}_{-0.032}$
&
$6.1$
\\
\end{tabular}
\caption{\label{tab:exprat}
Ratios of observed and predicted events in the Gallium source experiments,
the average ratio
$\overline{R}^{\text{HR}}$,
and the size of the Gallium Anomaly (GA)
for the different cross section models.
The predictions of all these models
assumed the Hampel and Remsberg (HR)
${}^{71}\text{Ge}$ half life.
}
\end{table*}

Table~\ref{tab:exprat} shows the
ratios of observed and predicted events in the Gallium source experiments
and
the average ratio
$\overline{R}^{\text{HR}}$
for the Ground State cross section model
and the models in Table~\eqref{tab:BGT}.
The ``HR'' superscript emphasizes that
the predictions of all these models
assumed the Hampel and Remsberg
${}^{71}\text{Ge}$ half life.

\begin{table*}
\setlength{\tabcolsep}{5pt}
\centering
\begin{tabular}{l|ccc|ccc}
&
\multicolumn{3}{c|}{$^{51}\text{Cr}$}
&
\multicolumn{3}{c|}{$^{37}\text{Ar}$}
\\
\cline{2-4}
\cline{5-7}
Model
&
$\sigma_{\text{mod}}$
&
$\overline{\sigma}_{\text{mod}} / \overline{\sigma}_{\text{gs}}$
&
$\Delta\eta_{\text{mod}}$
&
$\sigma_{\text{mod}}$
&
$\overline{\sigma}_{\text{mod}} / \overline{\sigma}_{\text{gs}}$
&
$\Delta\eta_{\text{mod}}$
\\
\hline
Ground State~\cite{Semenov:2020xea}
&
$5.539 \pm 0.019$
&
$1$
&
$0.003$
&
$6.625 \pm 0.023$
&
$1$
&
$0.003$
\\
Bahcall~\cite{Bahcall:1997eg}
&
$5.81 \pm 0.16$
&
$1.05$
&
$0.028$
&
$7.00 \pm 0.21$
&
$1.06$
&
$0.030$
\\
Haxton~\cite{Haxton:1998uc}
&
$6.39 \pm 0.65$
&
$1.15$
&
$0.102$
&
$7.72 \pm 0.81$
&
$1.17$
&
$0.105$
\\
Frekers~\cite{Frekers:2015wga}
&
$5.92 \pm 0.11$
&
$1.07$
&
$0.019$
&
$7.15 \pm 0.14$
&
$1.08$
&
$0.020$
\\
Kostensalo~\cite{Kostensalo:2019vmv}
&
$5.67 \pm 0.06$
&
$1.02$
&
$0.011$
&
$6.80 \pm 0.08$
&
$1.03$
&
$0.012$
\\
Semenov~\cite{Semenov:2020xea}
&
$5.94 \pm 0.12$
&
$1.07$
&
$0.020$
&
$7.17 \pm 0.15$
&
$1.08$
&
$0.021$
\\
\hline
\end{tabular}
\caption{\label{tab:cs}
Values of the cross sections in the different cross section models in units of $10^{-45}$~cm$^2$.
Also shown are the values of the ratio
$\overline{\sigma}_{\text{mod}} / \overline{\sigma}_{\text{gs}}$
and the uncertainty
$\Delta\eta_{\text{mod}}$
needed in Eqs.~\eqref{eq:chi2xi} and~\eqref{xitilde}.
}
\end{table*}

The results for the average ratios and the corresponding significance
of the Gallium Anomaly are slightly different from those obtained in Ref.~\cite{Giunti:2022btk},
because here we consider the improved $\chi^2$ function
\begin{equation}
\chi^2
=
\min_{\eta}
\left[
\sum_{\text{exp}}
\left( \dfrac{ R_{\text{exp}} - \eta \overline{R} }{ \Delta R_{\text{exp}} } \right)^2
+
\chi^2_{\eta}
\right]
,
\label{eq:chi2}
\end{equation}
with
$ R_{\text{exp}} \pm \Delta R_{\text{exp}} $
given in Table~\ref{tab:exprat}
for
$
\text{exp}
\in
\{
\text{GALLEX-1},
\text{GALLEX-2},
\text{SAGE-Cr},
\text{SAGE-Ar},
\text{BEST-in},
\text{BEST-out}
\}
$
and the different cross section models under the assumption of the
HR ${}^{71}\text{Ge}$ half life.
The pull factor $\eta$ takes into account the correlated uncertainties
of the ${}^{51}\text{Cr}$ and ${}^{37}\text{Ar}$ cross sections.
In the contribution $\chi^2_{\eta}$
of the pull factor $\eta$, we took into account
the lower limit of the cross section represented by the
Ground State model:
\begin{equation}
\chi^2_{\eta}
=
\left\{
\begin{array}{lcl} \displaystyle
\left(
\dfrac{ 1 - \eta }{ \Delta\eta_{\text{mod}} }
\right)^2
& \displaystyle
\text{for}
& \displaystyle
\eta \geq \tilde\eta
,
\\ \displaystyle
\left(
\dfrac
{ 1 - \eta \overline{\sigma}_{\text{mod}} / \overline{\sigma}_{\text{gs}} }
{ \Delta\eta_{\text{gs}} }
\right)^2
& \displaystyle
\text{for}
& \displaystyle
\eta < \tilde\eta
,
\end{array}
\right.
\label{eq:chi2xi}
\end{equation}
with
\begin{equation}
\tilde\eta
=
\dfrac
{ \Delta\eta_{\text{mod}} - \Delta\eta_{\text{gs}} }
{ \Delta\eta_{\text{mod}} \,
\overline{\sigma}_{\text{mod}} / \overline{\sigma}_{\text{gs}}
- \Delta\eta_{\text{gs}} }
.
\label{xitilde}
\end{equation}
Here $\overline{\sigma}_{\text{mod}}$ is the central value of the
model cross section under consideration and
$\overline{\sigma}_{\text{gs}}$
is the central value of the Ground State cross section.
The quantities $\Delta\eta_{\text{mod}}$
and
$\Delta\eta_{\text{gs}}$ are the relative uncertainties
of the model cross section under consideration
and of the Ground State cross section.
The values of
$\overline{\sigma}_{\text{mod}} / \overline{\sigma}_{\text{gs}}$
and
$\Delta\eta_{\text{mod}}$
are shown in Table~\ref{tab:cs}.
One can see that there is a slight difference of the values of these quantities
for ${}^{51}\text{Cr}$ and ${}^{37}\text{Ar}$ source experiments.
Since this difference cannot be taken into account in the $\chi^2$ function
and the result of the analysis of the Gallium data is dominated
by the ${}^{51}\text{Cr}$ source experiments,
we neglect it and we consider the values of
$\overline{\sigma}_{\text{mod}} / \overline{\sigma}_{\text{gs}}$
and
$\Delta\eta_{\text{mod}}$
for the ${}^{51}\text{Cr}$ source experiments.

\begin{figure}
\centering
\includegraphics[width=0.5\linewidth]{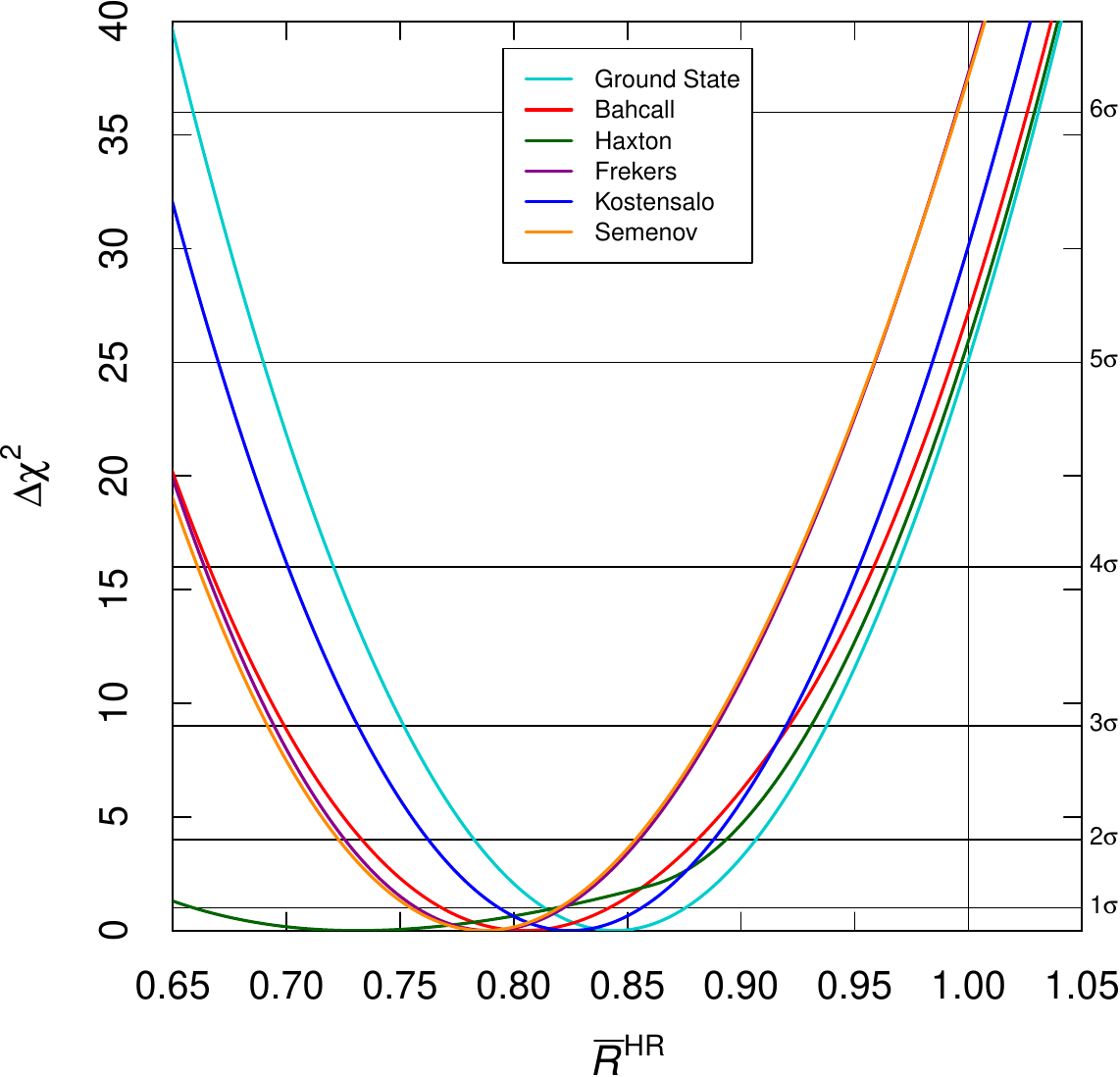}
\caption{\label{fig:chi2Rav}
$\Delta\chi^2 = \chi^2 - \chi^2_{\text{min}}$
as a function of the average ratio $\overline{R}^{\text{HR}}$
for the cross section models in Table~\ref{tab:exprat}.
}
\end{figure}

Figure~\ref{fig:chi2Rav} shows the value of
$\Delta\chi^2 = \chi^2 - \chi^2_{\text{min}}$
as a function of the average ratio $\overline{R}^{\text{HR}}$
for the different cross section models,
from which we calculated the best-fit values and the $1\sigma$ uncertainties of
$\overline{R}^{\text{HR}}$,
and the size of the Gallium Anomaly
reported in the last two columns of Table~\ref{tab:exprat}.
One can see that for all cross section models
the value of $\Delta\chi^2$ is larger than that
of the Ground State model
when $\overline{R}^{\text{HR}}$ is larger than the best fit value.
This is assured by the contribution of $\chi^2_{\eta}$
to the total $\chi^2$.
The effect is particularly clear in the case of the Haxton model,
for which one can see that the corresponding $\Delta\chi^2$
in Fig.~\ref{fig:chi2Rav}
deviates from a parabola at $\overline{R}^{\text{HR}} \simeq 0.87$.
This behaviour is required for
$\overline{R}^{\text{HR}}$ larger than the best fit value,
because a large value of $\overline{R}^{\text{HR}}$
corresponds to a small cross section,
which is bounded by the Ground State lower limit.

A consequence of the $\chi^2$ function that we have adopted
is that the sizes of the Gallium Anomaly obtained with all the cross section models
with transitions to the excited states of ${}^{71}\text{Ge}$
are larger than that obtained with the Ground State model,
as shown by the last column in Table~\ref{tab:exprat}.
This is correct and expected
from the lower bound on the cross section represented by
the Ground State model.

Comparing the sizes of the Gallium Anomaly
in Table~\ref{tab:exprat}
with those presented in Table~1 of Ref.~\cite{Giunti:2022btk}
one can see that there are slight variations
which are due to the improvement of the $\chi^2$ function.
The size of the Gallium Anomaly obtained here with the
Haxton model is larger than that obtained
in Ref.~\cite{Giunti:2022btk}
because of the Ground State lower bound on the cross section
discussed above, which was not implemented in Ref.~\cite{Giunti:2022btk}.
The sizes of the Gallium Anomaly obtained here with the
Bahcall, Frekers, Kostensalo, and Semenov models
are slightly smaller than those obtained
in Ref.~\cite{Giunti:2022btk},
because the $\chi^2$ function used in Ref.~\cite{Giunti:2022btk}
is the usual one with the correlated uncertainty taken into account
in a covariance matrix.
This $\chi^2$ function suffers of the problem called
``Peelle's Pertinent Puzzle'' (PPP)
discussed in Ref.~\cite{Giunti:2021kab},
which leads to best-fit values of
$\overline{R}^{\text{HR}}$
which are smaller than that obtained
from the weighted average of the experimental values
and could be smaller than all or most of the
experimental values if the correlated uncertainty is large.
The $\chi^2$ function~\eqref{eq:chi2} that we adopted here
avoids this problem
(see the clear discussion in Ref.~\cite{DAgostini:1993arp}).
Therefore,
with the Haxton model, which has the largest correlated uncertainty
(see Table~\ref{tab:BGT}),
we obtain a best-fit value of 0.731 for $\overline{R}^{\text{HR}}$,
which is significantly larger than
the 0.703 obtained in Ref.~\cite{Giunti:2022btk}.
For the Ground State, Bahcall, Frekers, Kostensalo, and Semenov models,
which have smaller correlated uncertainties,
we obtain best-fit values of $\overline{R}^{\text{HR}}$
which are only slightly larger than those obtained in Ref.~\cite{Giunti:2022btk}.
This effect causes the slight decrease
of the sizes of the Gallium Anomaly obtained here with the
Bahcall, Frekers, Kostensalo, and Semenov models
with respect to those obtained in Ref.~\cite{Giunti:2022btk}.
The size of the Gallium Anomaly
obtained here with the Haxton model
is not smaller than that obtained in Ref.~\cite{Giunti:2022btk}
in spite of the increase of the best-fit values of $\overline{R}^{\text{HR}}$,
because of the Ground State lower bound on the cross section
discussed above.

\begin{figure*}
\centering
\subfigure[]{ \label{fig:ge71-groundstate}
\includegraphics[width=0.31\linewidth]{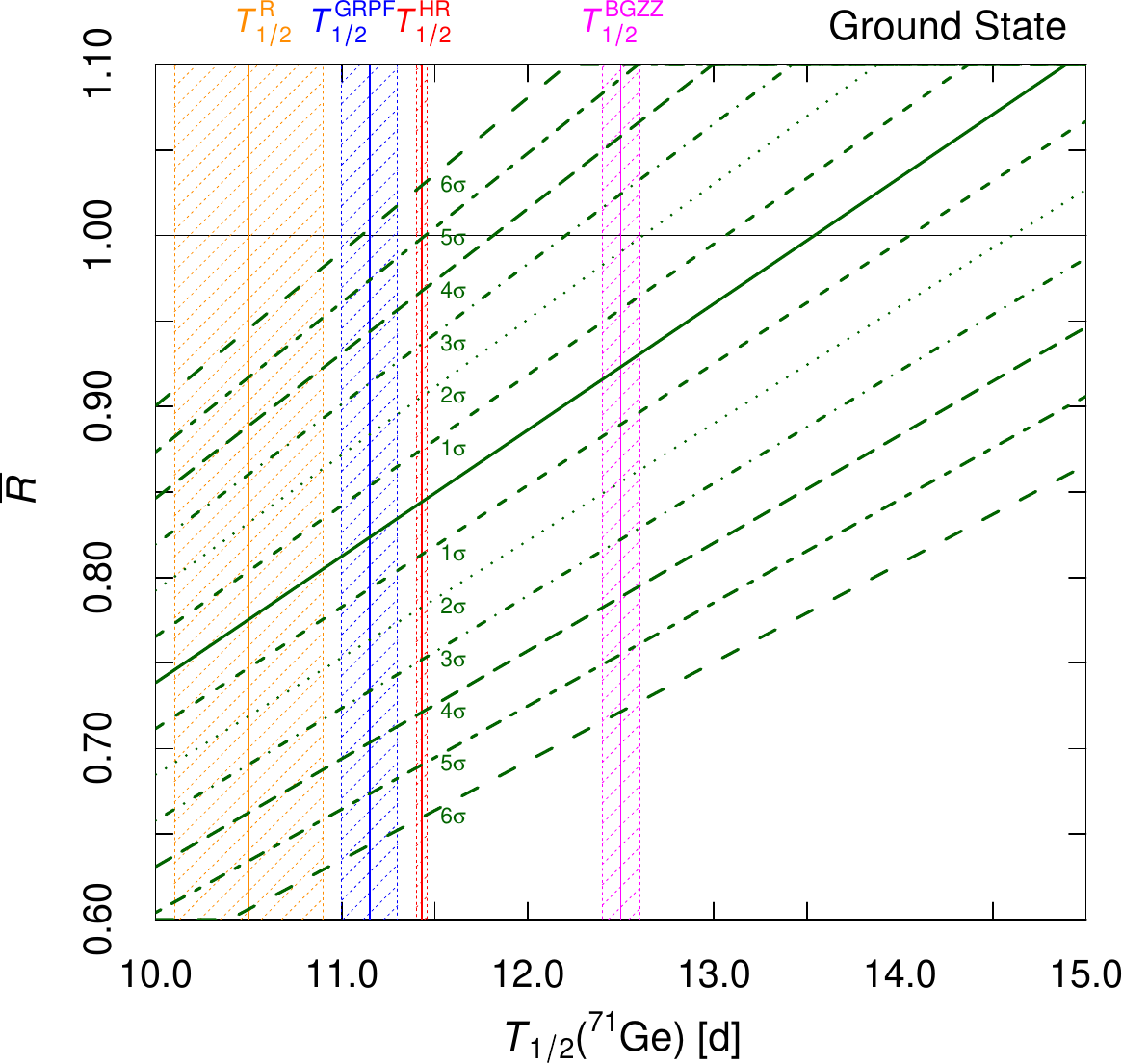}
}
\subfigure[]{ \label{fig:ge71-bahcall}
\includegraphics[width=0.31\linewidth]{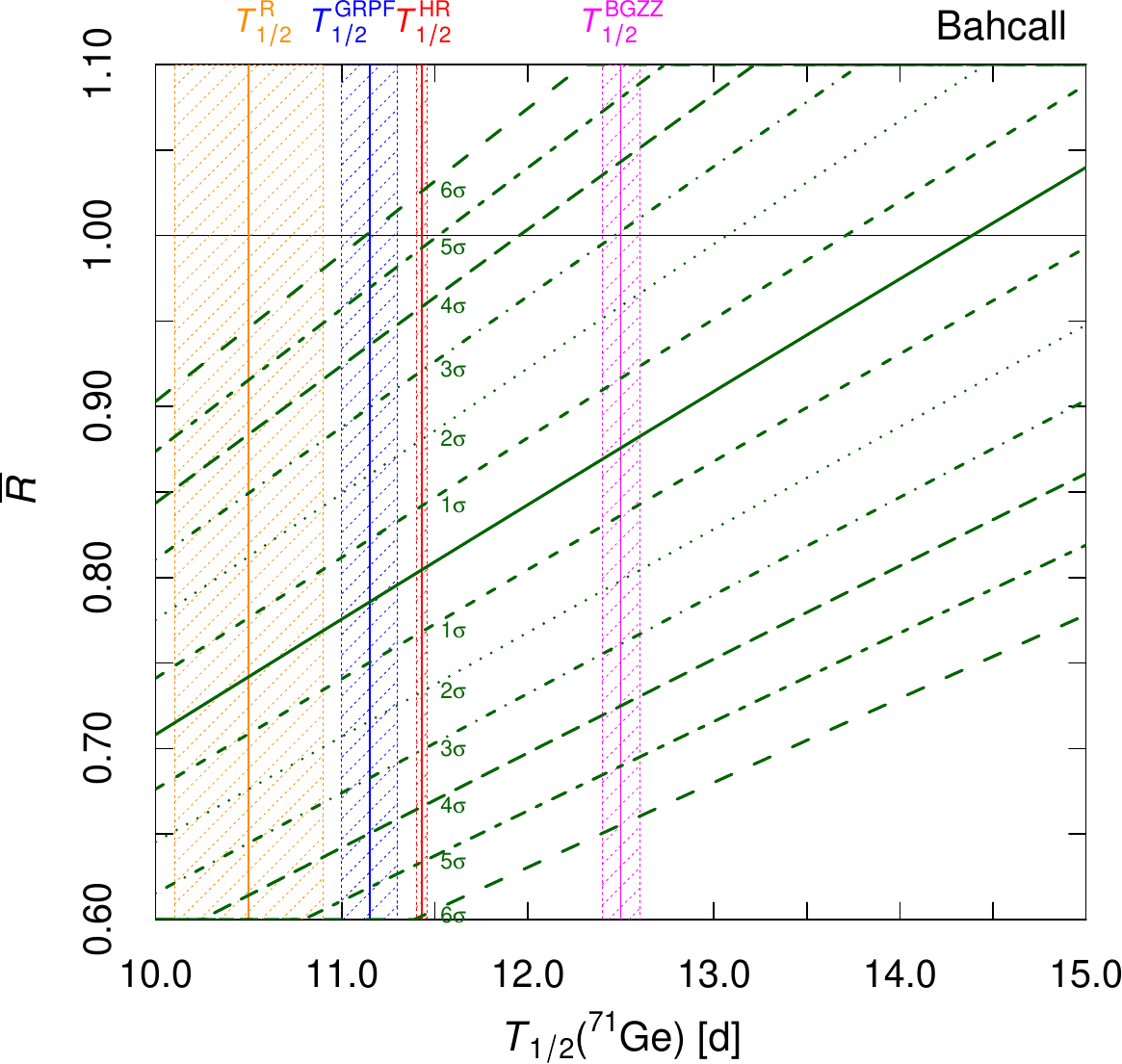}
}
\subfigure[]{ \label{fig:ge71-haxton}
\includegraphics[width=0.31\linewidth]{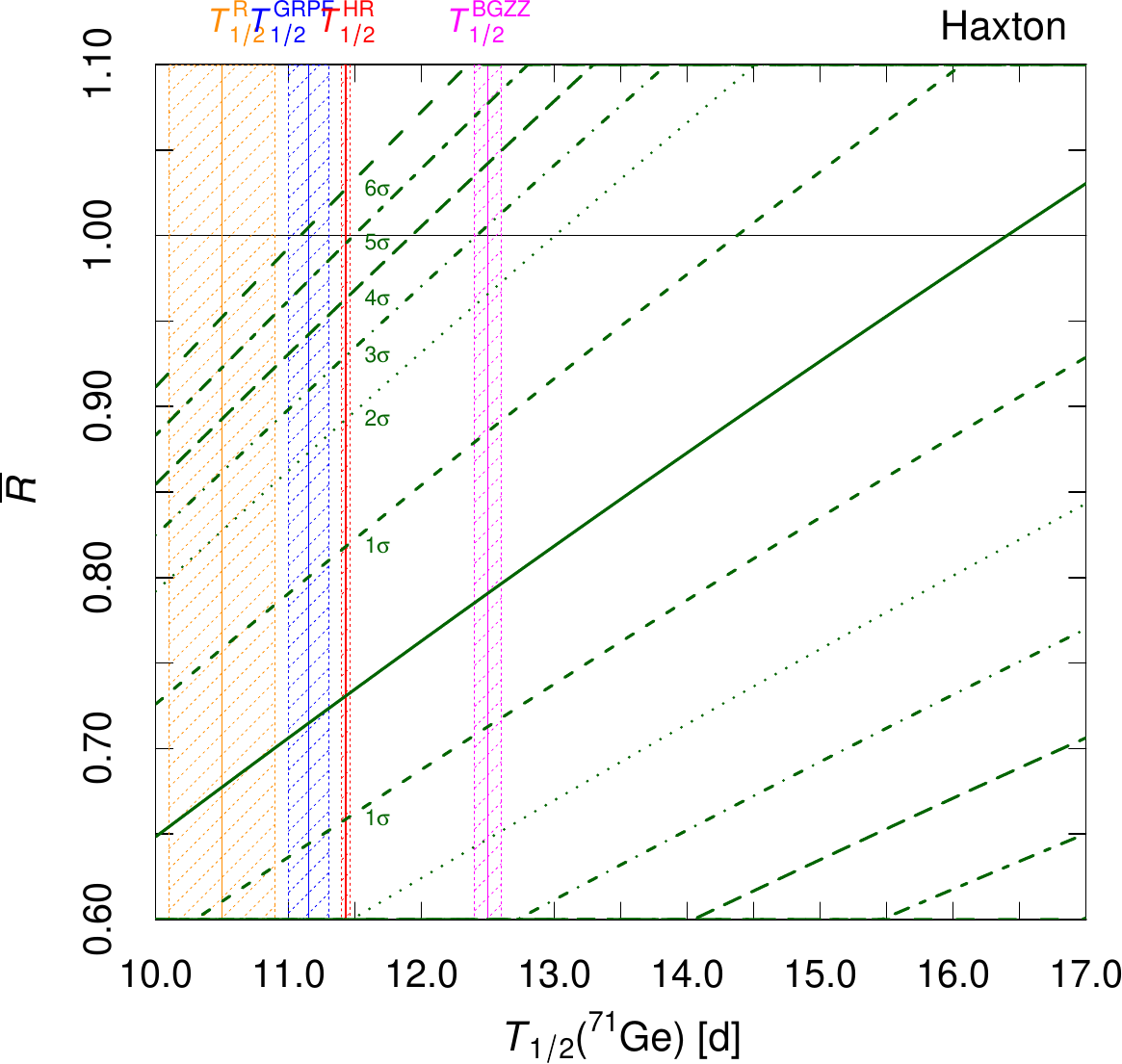}
}
\\
\subfigure[]{ \label{fig:ge71-frekers}
\includegraphics[width=0.31\linewidth]{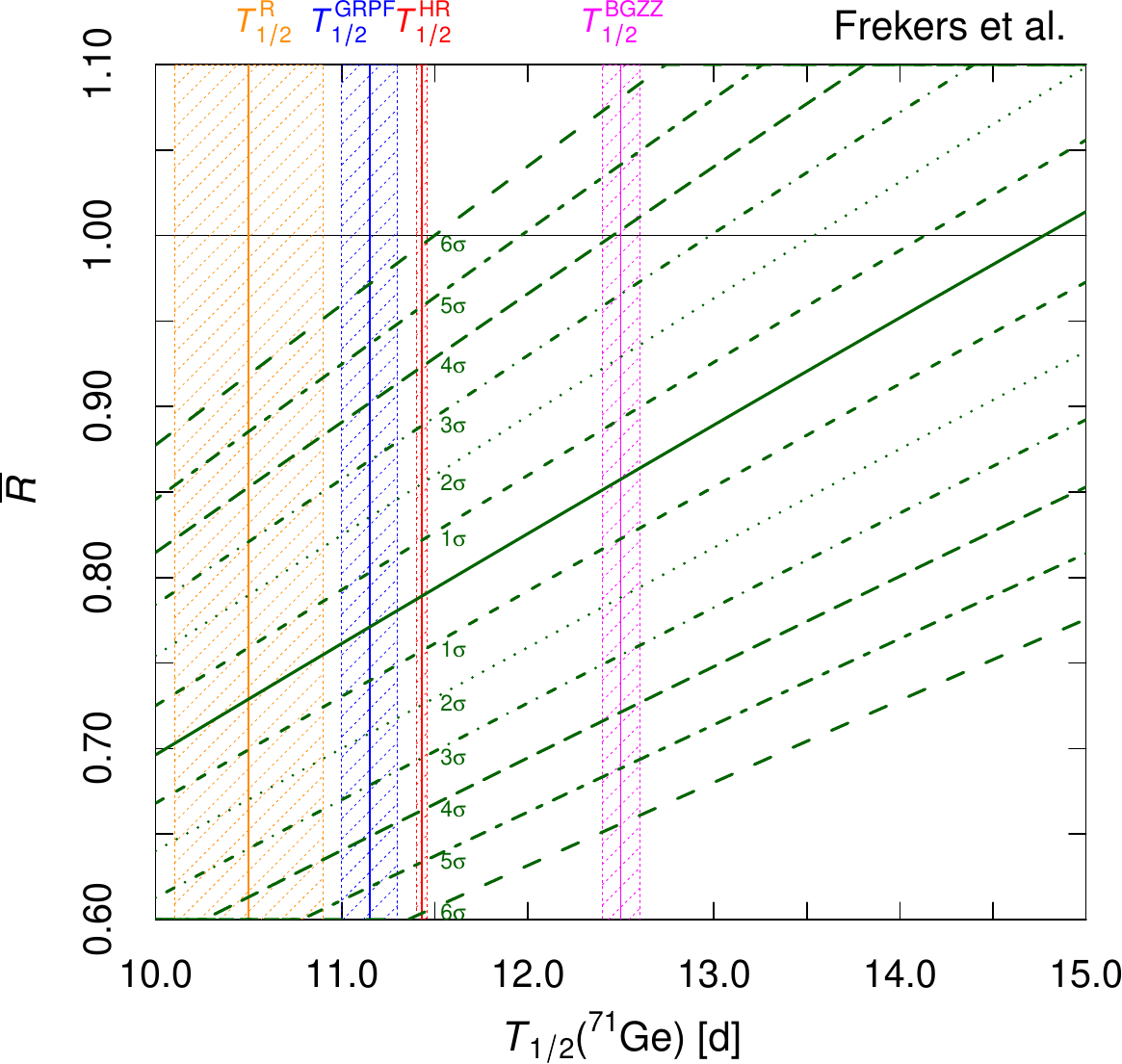}
}
\subfigure[]{ \label{fig:ge71-kostensalo}
\includegraphics[width=0.31\linewidth]{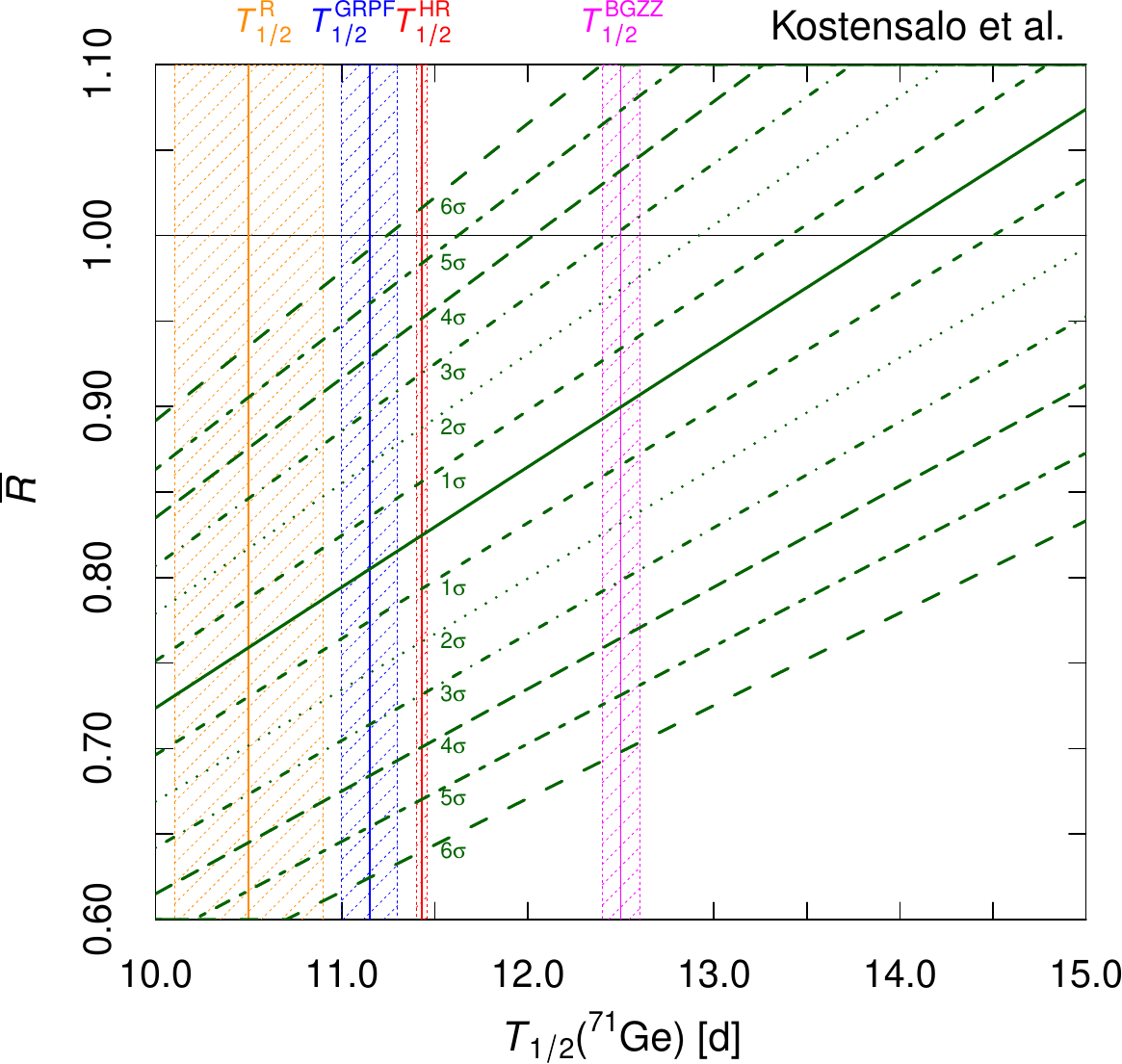}
}
\subfigure[]{ \label{fig:ge71-semenov}
\includegraphics[width=0.31\linewidth]{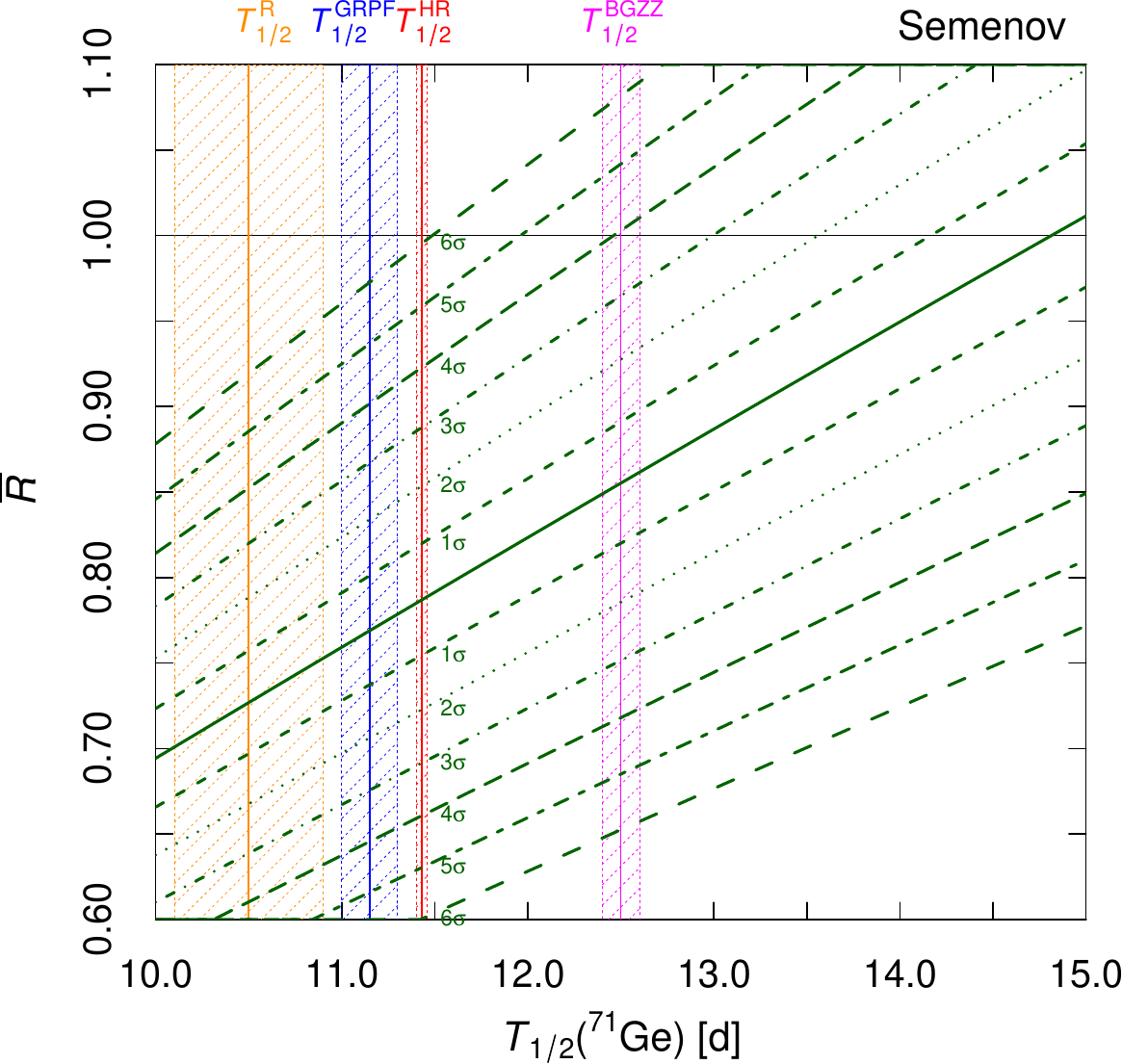}
}
\caption{\label{fig:ge71-models}
Dependence of the average ratio $\overline{R}$
on $T_{1/2}({}^{71}\text{Ge})$
for the cross section models in Table~\ref{tab:exprat}.
The central solid green line in the best-fit value of $\overline{R}$.
The other green lines delimit the $1\sigma$, \ldots, $6\sigma$
bands as indicated by the corresponding labels.
The vertical bands show the measurements of $T_{1/2}({}^{71}\text{Ge})$
in Eqs.~\eqref{eq:BGZZ}--\eqref{eq:HR}.
}
\end{figure*}

\section{The ${}^{71}\text{Ge}$ half life}
\label{sec:lifetime}

The ${}^{71}\text{Ge}$ half life measurement in Eq.~\eqref{eq:HR}
has never been questioned so far,
but an inspection of Ref.~\cite{Hampel:1985zz} reveals two puzzling aspects:
\begin{enumerate}
\item
The measurement was done in 1985.
It is puzzling that this measurement,
which is crucial for the Gallium Anomaly
has not been checked with a modern technique and apparatus.
\item
The other previous measurements of $T_{1/2}({}^{71}\text{Ge})$
in Eqs.~\eqref{eq:BGZZ}--\eqref{eq:GRPF}
gave different values, albeit with larger uncertainties.
\end{enumerate}
In view of these considerations,
it is worth to study what happens to the Gallium Anomaly
by changing $T_{1/2}({}^{71}\text{Ge})$.
Since $ft_{1/2}({}^{71}\text{Ge})$
is proportional to $T_{1/2}({}^{71}\text{Ge})$,
from Eq.~\eqref{eq:sigmags} it is clear that
\begin{equation}
\sigma_{\text{gs}}
\propto
\text{BGT}_{\text{gs}}
\propto
1 / T_{1/2}({}^{71}\text{Ge})
.
\label{eq:siggsprop}
\end{equation}
Then,
from Eq.~\eqref{eq:cs},
the full detection cross section can be written as
\begin{equation}
\sigma_{\text{tot}}
=
\sigma_{\text{gs}}^{\text{HR}}
\left(
\dfrac{ T_{1/2}^{\text{HR}}({}^{71}\text{Ge}) }{ T_{1/2}({}^{71}\text{Ge}) }
+
\xi_{5/2^{-}}
\dfrac{\text{BGT}_{5/2^{-}}}{\text{BGT}_{\text{gs}}^{\text{HR}}}
+
\xi_{3/2^{-}}
\dfrac{\text{BGT}_{3/2^{-}}}{\text{BGT}_{\text{gs}}^{\text{HR}}}
+
\xi_{5/2^{+}}
\dfrac{\text{BGT}_{5/2^{+}}}{\text{BGT}_{\text{gs}}^{\text{HR}}}
\right)
.
\label{eq:csnew}
\end{equation}

The Gallium Anomaly has been so far evaluated assuming the Hampel and Remsberg
${}^{71}\text{Ge}$ half life.
Table~\ref{tab:exprat} shows the
ratios of observed and predicted events in the Gallium source experiments
and
the average ratio
$\overline{R}^{\text{HR}}$
for the different cross section models
discussed in Section~\ref{sec:detection}.
The ``HR'' superscript emphasizes that
the predictions of all these models
assumed the Hampel and Remsberg
${}^{71}\text{Ge}$ half life.

If we want to consider another ${}^{71}\text{Ge}$ half life,
the corresponding ratio of measured and predicted ${}^{71}\text{Ge}$ production rates
is given by
\begin{equation}
R
=
R^{\text{HR}}
\,
\dfrac{
1
+
\xi_{5/2^{-}}
\dfrac{\text{BGT}_{5/2^{-}}}{\text{BGT}_{\text{gs}}^{\text{HR}}}
+
\xi_{3/2^{-}}
\dfrac{\text{BGT}_{3/2^{-}}}{\text{BGT}_{\text{gs}}^{\text{HR}}}
+
\xi_{5/2^{+}}
\dfrac{\text{BGT}_{5/2^{+}}}{\text{BGT}_{\text{gs}}^{\text{HR}}}
}{
\dfrac{ T_{1/2}^{\text{HR}}({}^{71}\text{Ge}) }{ T_{1/2}({}^{71}\text{Ge}) }
+
\xi_{5/2^{-}}
\dfrac{\text{BGT}_{5/2^{-}}}{\text{BGT}_{\text{gs}}^{\text{HR}}}
+
\xi_{3/2^{-}}
\dfrac{\text{BGT}_{3/2^{-}}}{\text{BGT}_{\text{gs}}^{\text{HR}}}
+
\xi_{5/2^{+}}
\dfrac{\text{BGT}_{5/2^{+}}}{\text{BGT}_{\text{gs}}^{\text{HR}}}
}
.
\label{eq:newrat}
\end{equation}

In order to evaluate the ratio $R$ corresponding to some value of $T_{1/2}({}^{71}\text{Ge})$
for a cross section model,
we need to know the corresponding BGT ratios in Eq.~\eqref{eq:newrat}.
These values are given in Table~\ref{tab:BGT} for the cross section models in Table~\ref{tab:exprat}
(except for the Ground State model, for which they are absent).

The contributions of the transitions to the excited states of
${}^{71}\text{Ge}$
complicate the calculation of the average ratio $\overline{R}$,
because these contributions are different for the
${}^{51}\text{Cr}$ and ${}^{37}\text{Ar}$
sources.
Therefore,
the correction~\eqref{eq:newrat}
must be applied to the ratio of each source experiment in Table~\ref{tab:exprat}
and the results must be used to calculate the average ratio
(the outcomes are presented in Section~\ref{sec:dependence}).
This complication is not necessary for the Ground State model.

In the Ground State model
the contributions of the transitions to the excited states of ${}^{71}\text{Ge}$
are assumed to be negligible.
It is an extreme possibility that is justified
by the uncertainties of the cross sections
to the excited states of ${}^{71}\text{Ge}$,
which depend on the methods and assumptions of the different models.

In the Ground state model, the relation~\eqref{eq:newrat}
becomes simply
\begin{equation}
R_{\text{Ground State}}
=
R_{\text{Ground State}}^{\text{HR}}
\,
\dfrac{ T_{1/2}({}^{71}\text{Ge}) }
      { T_{1/2}^{\text{HR}}({}^{71}\text{Ge}) }
.
\label{eq:newratioGS}
\end{equation}
Since it is a common rescaling of all the ratios of the source experiments,
it can be applied directly to the average ratio.

In particular we can find which value of $T_{1/2}({}^{71}\text{Ge})$
is required to cancel the Gallium Anomaly
by considering $\overline{R}_{\text{Ground State}}=1$:
\begin{equation}
T_{1/2}({}^{71}\text{Ge})
=
\dfrac{ T_{1/2}^{\text{HR}}({}^{71}\text{Ge}) }
      { \overline{R}_{\text{Ground State}}^{\text{HR}} }
=
13.5 \pm 0.5 \, \text{d}
,
\label{eq:TGS}
\end{equation}
using the value in
Eq.~\eqref{eq:HR} and Table~\ref{tab:exprat}.
This value of $T_{1/2}({}^{71}\text{Ge})$
is rather large.
It is larger than all the measurements
in Eqs.~\eqref{eq:BGZZ}--\eqref{eq:HR}.
However,
taking into account the oldness of these measurements,
it may be not completely unrealistic.

We can also perform a simple evaluation of
the value of the average ratio that would be obtained
by considering the largest measurement
of $T_{1/2}({}^{71}\text{Ge})$ in Eq.~\eqref{eq:BGZZ}:
\begin{equation}
\overline{R}_{\text{Ground State}}^{\text{BGZZ}}
=
\overline{R}_{\text{Ground State}}^{\text{HR}}
\,
\dfrac{ T_{1/2}^{\text{BGZZ}}({}^{71}\text{Ge}) }
      { T_{1/2}^{\text{HR}}({}^{71}\text{Ge}) }
=
0.924 \pm 0.035
.
\label{BisiRGS}
\end{equation}
Therefore,
in the case of the Ground State model,
considering the largest measurement
of $T_{1/2}({}^{71}\text{Ge})$ in Eq.~\eqref{eq:BGZZ}
reduces the Gallium Anomaly to about $2.2\sigma$,
which is much smaller than the $5.0\sigma$
obtained with the standard value of Hampel and Remsberg. 

\begin{figure}
\centering
\includegraphics[width=0.5\linewidth]{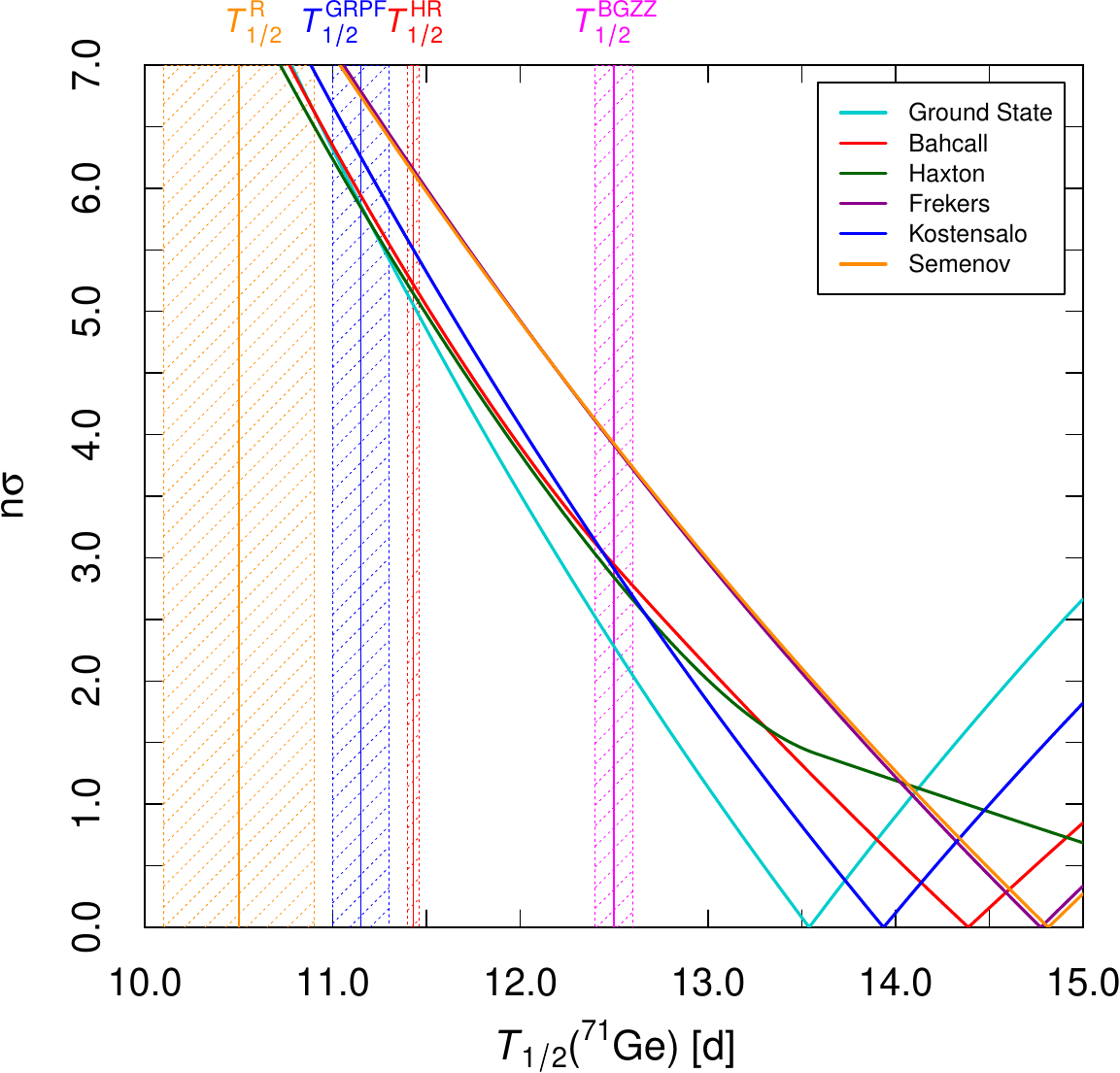}
\caption{\label{fig:sig}
Size of the Gallium Anomaly
as a function of $T_{1/2}({}^{71}\text{Ge})$
for the cross section models in Table~\ref{tab:exprat}.
The vertical bands show the measurements of $T_{1/2}({}^{71}\text{Ge})$
in Eqs.~\eqref{eq:BGZZ}--\eqref{eq:HR}.
}
\end{figure}

\section{Gallium Anomaly dependence on $T_{1/2}({}^{71}\text{Ge})$}
\label{sec:dependence}

Using Eq.~\eqref{eq:newrat},
we can determine the dependence
on $T_{1/2}({}^{71}\text{Ge})$
of the ratio
of measured and predicted ${}^{71}\text{Ge}$ production rates
for each source experiment
and we can calculate the dependence for the average ratio $\overline{R}$
obtained by minimizing the $\chi^2$ function in Eq.~\eqref{eq:chi2}.

Figure~\ref{fig:ge71-models}
shows the dependence of the average ratio $\overline{R}$
on $T_{1/2}({}^{71}\text{Ge})$
for the Ground state model and the cross section models in Table~\ref{tab:exprat},
with the uncertainties due to the experimental uncertainties and
the theoretical uncertainties of the Gamow-Teller strengths
in Table~\ref{tab:BGT}.
One can see that for the standard value $T_{1/2}^{\text{HR}}({}^{71}\text{Ge})$
the deviation of $\overline{R}$ from unity corresponds to the
size of the Gallium anomaly in the last column of Table.~\ref{tab:exprat}.
The smaller values of $T_{1/2}({}^{71}\text{Ge})$
obtained in the measurements in Eqs.~\eqref{eq:R} and~\eqref{eq:GRPF}
obviously lead to a larger Gallium Anomaly,
whereas the BGZZ value in Eq.~\eqref{eq:BGZZ}
reduces the Gallium Anomaly to a level of about $2\sigma$
for the Ground State cross section model,
at a level of about $3\sigma$ for the Bahcall, Haxton, and Kostensalo models,
and at a level of about $4\sigma$ for the Frekers and Semenov models.
As one can see from Fig.~\ref{fig:ge71-models},
larger values of $T_{1/2}({}^{71}\text{Ge})$ can cancel entirely
the Gallium Anomaly.

The dependence on
$T_{1/2}({}^{71}\text{Ge})$
of the size of the Gallium Anomaly
is shown in Fig.~\ref{fig:sig}
for the different cross section models.
One can see that $T_{1/2}({}^{71}\text{Ge}) \gtrsim 13.5 \, \text{d}$
is necessary in order to reduce the Gallium Anomaly below about
$2\sigma$
for all the cross section models.
In particular for the Frekers and Semenov models
which have the largest Gamow-Teller strengths
shown in Table~\ref{tab:BGT}.
For the Bahcall, Haxton, and Kostensalo models,
the Gallium Anomaly can be reduced below about $2\sigma$
with a more reasonable
$T_{1/2}({}^{71}\text{Ge}) \gtrsim 13 \, \text{d}$.
For the Ground State model it is sufficient to have
$T_{1/2}({}^{71}\text{Ge}) \gtrsim 12.5 \, \text{d}$.

\begin{figure*}
\centering
\subfigure[]{ \label{fig:see-gal-groundstate}
\includegraphics[width=0.31\linewidth]{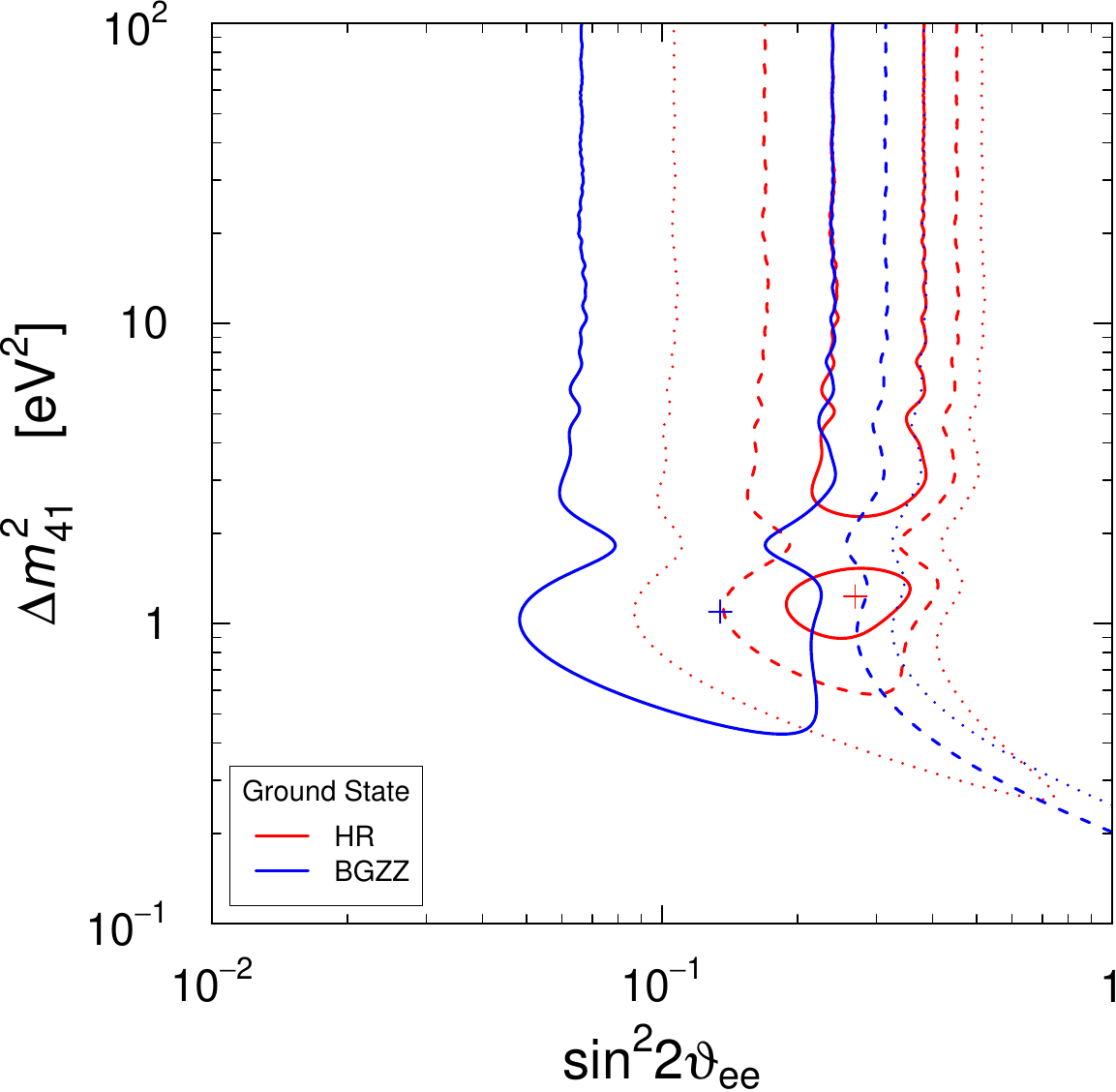}
}
\subfigure[]{ \label{fig:see-gal-bahcall}
\includegraphics[width=0.31\linewidth]{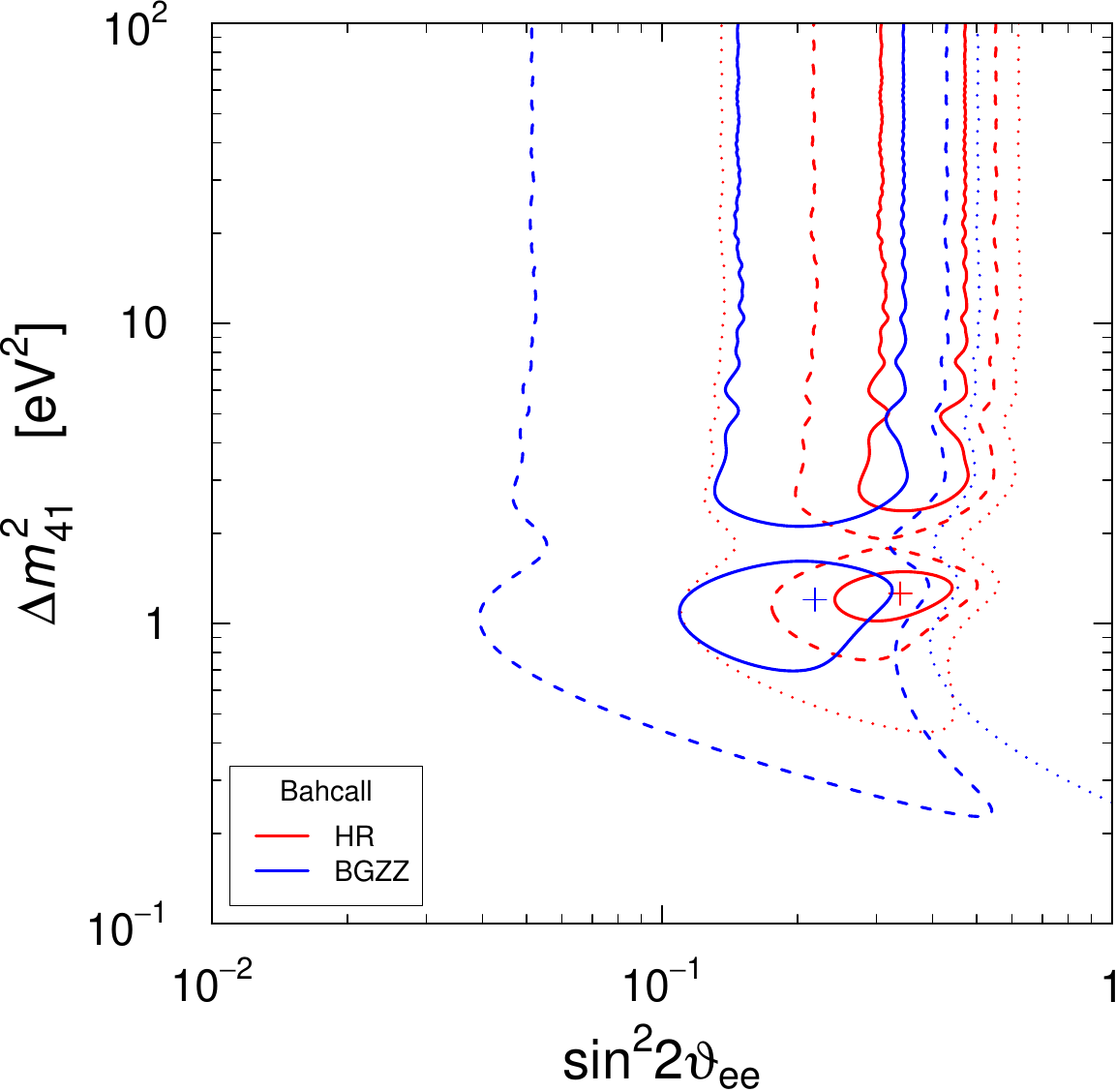}
}
\subfigure[]{ \label{fig:see-gal-haxton}
\includegraphics[width=0.31\linewidth]{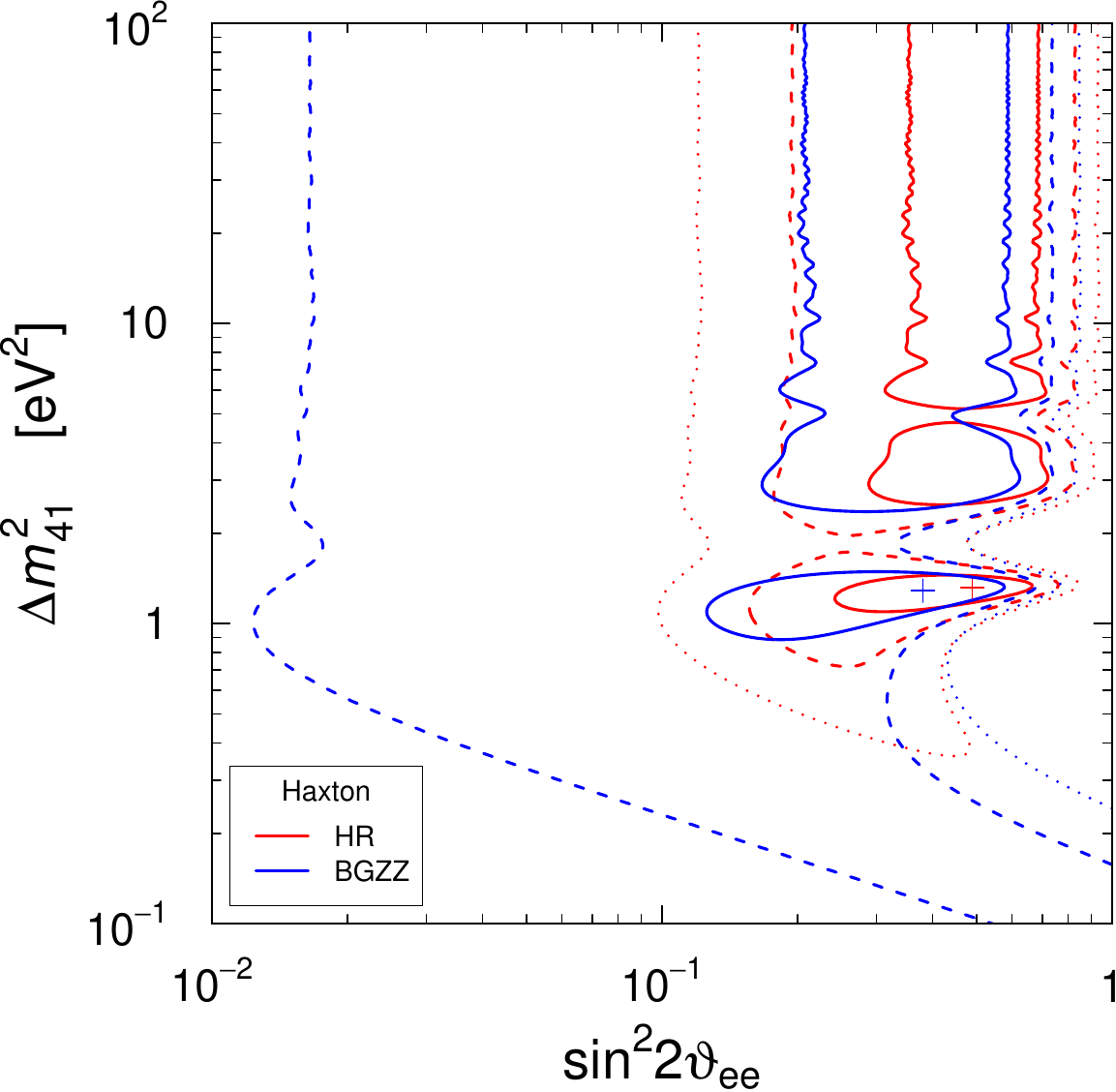}
}
\\
\subfigure[]{ \label{fig:see-gal-frekers}
\includegraphics[width=0.31\linewidth]{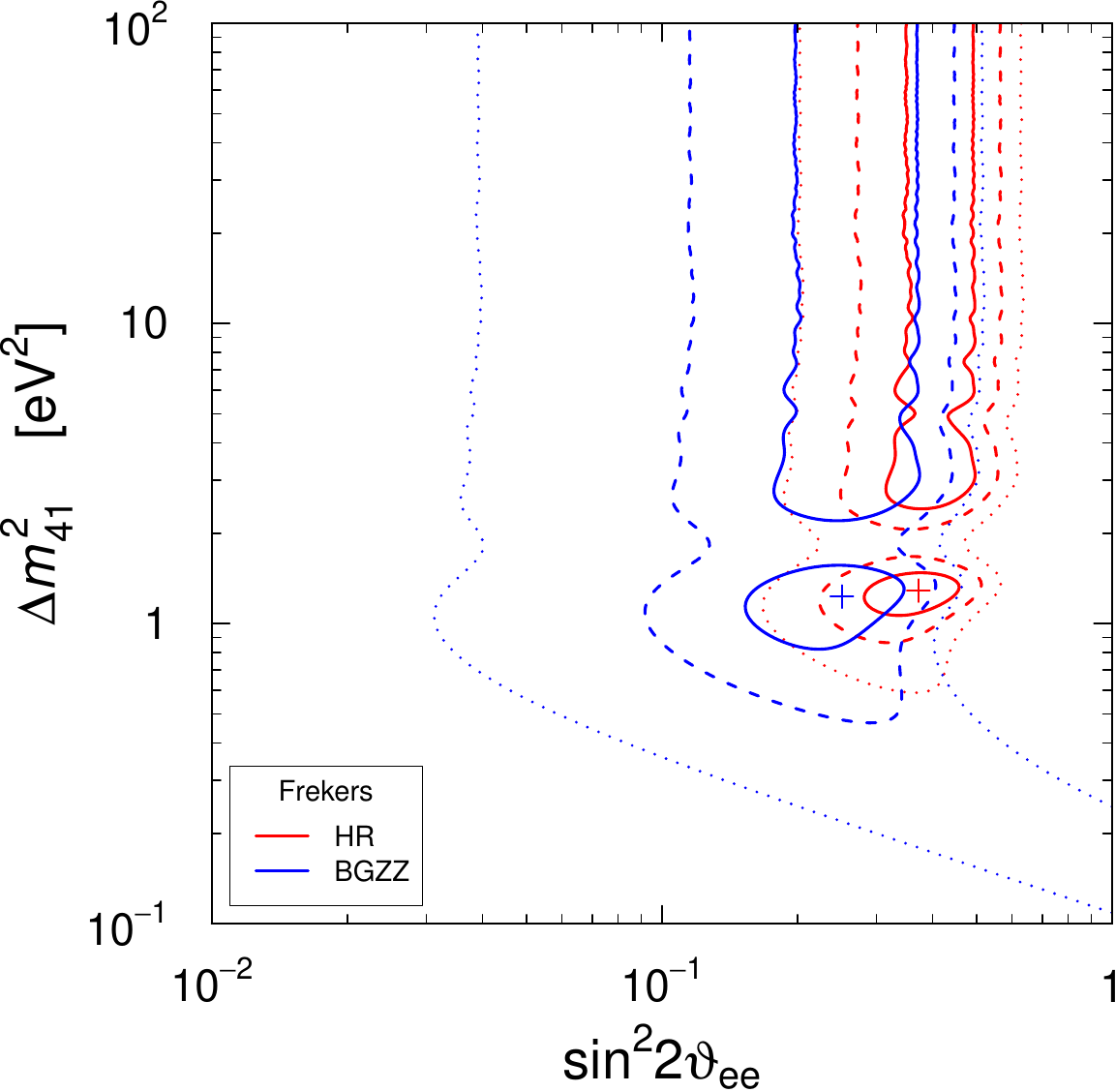}
}
\subfigure[]{ \label{fig:see-gal-kostensalo}
\includegraphics[width=0.31\linewidth]{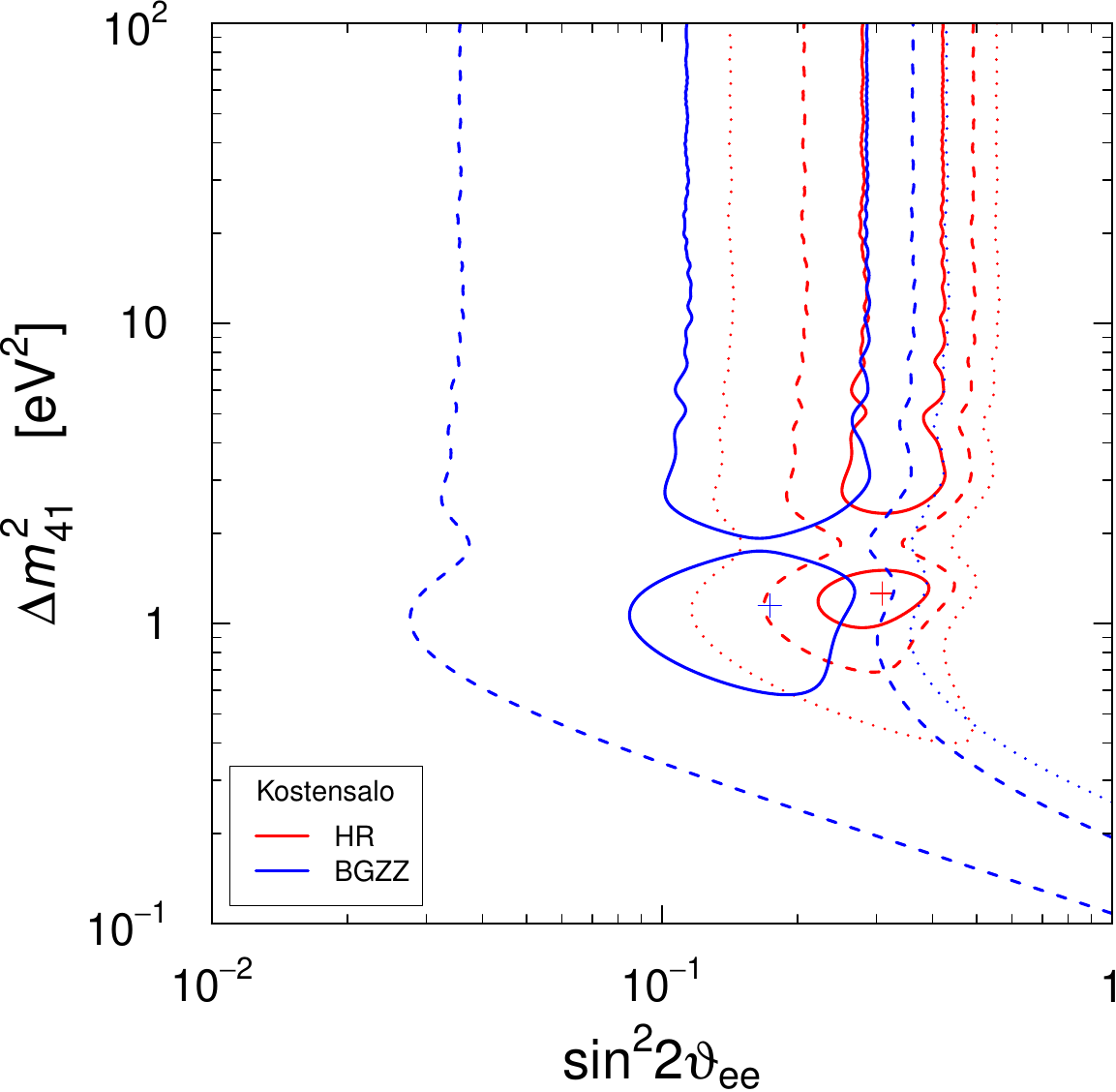}
}
\subfigure[]{ \label{fig:see-gal-semenov}
\includegraphics[width=0.31\linewidth]{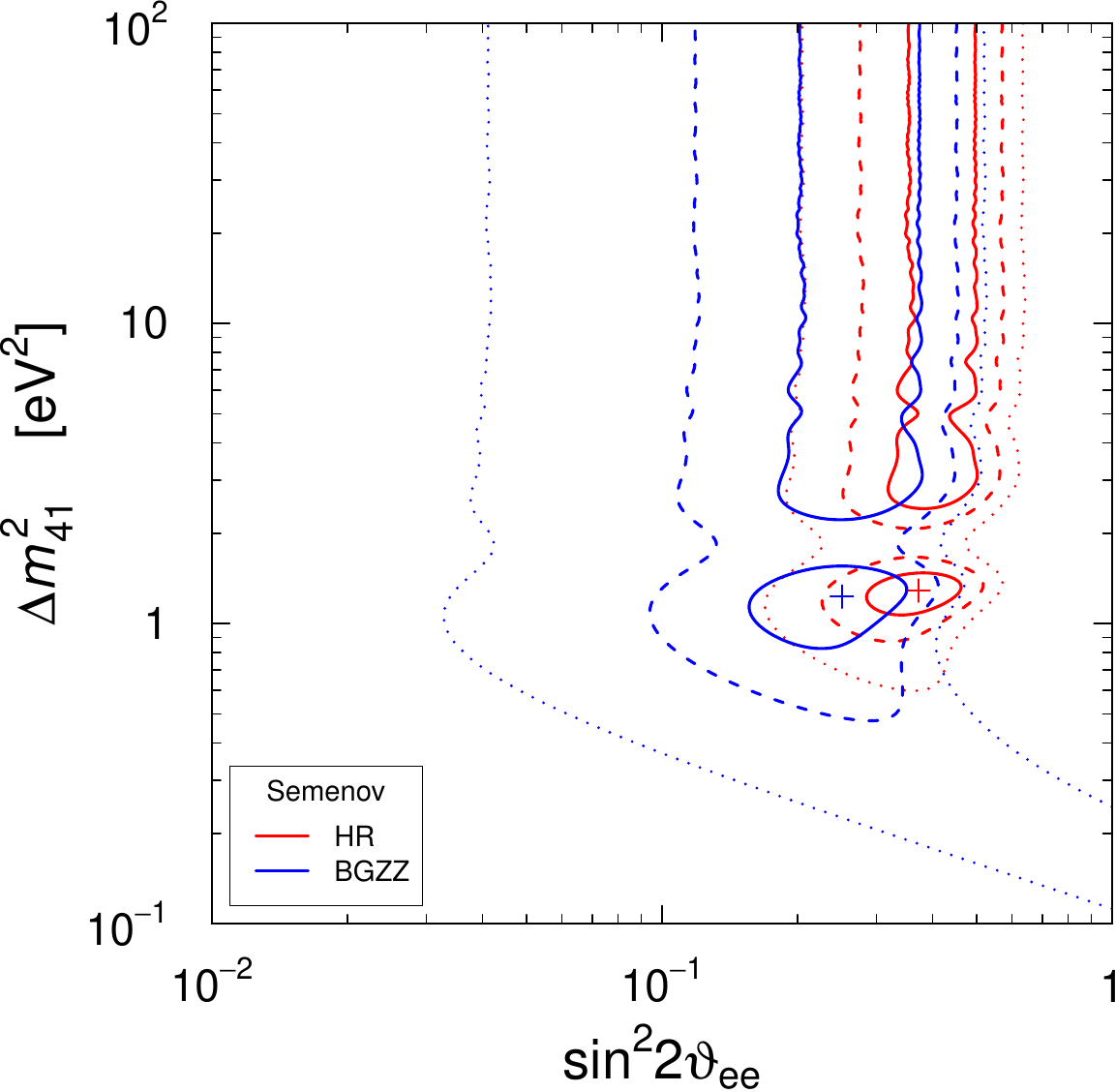}
}
\caption{\label{fig:see-gal-models}
Comparison of the
$1\sigma$ (solid),
$2\sigma$ (dashed), and
$3\sigma$ (dotted)
allowed regions in the
($\sin^2\!2\vartheta_{ee},\Delta{m}^2_{41}$)
plane obtained from the Gallium data
with the different cross section models
and the traditional Hampel and Remsberg (HR)
value \eqref{eq:HR} of $T_{1/2}({}^{71}\text{Ge})$
with those obtained with the larger
Bisi, Germagnoli, Zappa, and Zimmer (BGZZ)
value \eqref{eq:BGZZ} of $T_{1/2}({}^{71}\text{Ge})$.
}
\end{figure*}

\section{Short-baseline oscillations}
\label{sec:oscillations}

As explained in the introductory Section~\ref{sec:intro},
the Gallium Anomaly can be caused by
short-baseline oscillations due to active-sterile neutrino mixing.
However in the standard interpretation of the Gallium data
the required active-sterile neutrino mixing is large
and incompatible with other experimental bounds~\cite{Giunti:2022btk}.
We have shown in Section~\ref{sec:dependence} that the Gallium Anomaly
is reduced if the half life of ${}^{71}\text{Ge}$
is larger than the value of Eq.~\eqref{eq:HR} measured by Hampel and Remsberg
in 1985.
In a similar way,
an increase of $T_{1/2}({}^{71}\text{Ge})$
with respect to the Hampel and Remsberg value
changes the interpretation of the Gallium Anomaly
as due to short-baseline oscillations
by decreasing the required active-sterile neutrino mixing.

As an illustration of this effect,
we consider the BGZZ value of $T_{1/2}({}^{71}\text{Ge})$
in Eq.~\eqref{eq:BGZZ}.
Figure~\ref{fig:see-gal-models}
shows a comparison of the allowed regions
in the
($\sin^2\!2\vartheta_{ee},\Delta{m}^2_{41}$)
plane obtained with
$T_{1/2}^{\text{HR}}({}^{71}\text{Ge})$
and
$T_{1/2}^{\text{BGZZ}}({}^{71}\text{Ge})$
for the different detection cross section models.
One can see that for all the models
the adoption of $T_{1/2}^{\text{BGZZ}}({}^{71}\text{Ge})$
leads to a substantial shift of the allowed regions towards small values
of the effective active-sterile mixing angle
$\vartheta_{ee}$.

For the Ground State cross section model,
the adoption of $T_{1/2}^{\text{BGZZ}}({}^{71}\text{Ge})$
gives only $1\sigma$ regions which limit
$\sin^2\!2\vartheta_{ee}$ from below,
whereas at $2\sigma$ and $3\sigma$
there are only upper limits for $\sin^2\!2\vartheta_{ee}$
at $ \Delta{m}^2_{41} \gtrsim 0.2 \, \text{eV}^2 $.
Therefore, for the Ground State cross section model the absence of
short-baseline oscillations and active-sterile mixing
is allowed at $2\sigma$.

For the Bahcall, Haxton, and Kostensalo cross section models,
$\sin^2\!2\vartheta_{ee}$ is bounded from below
at $2\sigma$
if $T_{1/2}^{\text{BGZZ}}({}^{71}\text{Ge})$ is assumed.

Only for the Frekers and Semenov cross section models,
which have the largest Gamow-Teller strengths
shown in Table~\ref{tab:BGT},
the adoption of $T_{1/2}^{\text{BGZZ}}({}^{71}\text{Ge})$
leads to $3\sigma$ lower limits for $\sin^2\!2\vartheta_{ee}$.
However, for
$ \Delta{m}^2_{41} \gtrsim 1 \, \text{eV}^2 $
the $3\sigma$ lower limits lie at $\sin^2\!2\vartheta_{ee} \approx 0.04 $,
which is below or of the same order
of the upper bounds on $\sin^2\!2\vartheta_{ee}$
obtained in Ref.~\cite{Giunti:2022btk} from other experimental data.

Therefore,
the adoption of $T_{1/2}^{\text{BGZZ}}({}^{71}\text{Ge})$
leads to a strong reduction of the tension between the Gallium data and other data
which has been discussed in Ref.~\cite{Giunti:2022btk}.

\section{Summary and conclusions}
\label{sec:conclusions}

In this paper we have discussed in detail the dependence of the Gallium Anomaly on the detection cross section.
We have considered all the existing cross section models.
In Section~\ref{sec:detection} we have presented the results of an improved analysis of the data of the Gallium source experiments
which takes into account the lower bound on the cross section in each model
given by the transition from the ground state of
${}^{71}\text{Ga}$ to the ground state of ${}^{71}\text{Ge}$.
We have found that the size of the Gallium Anomaly is larger than about $5\sigma$ for all the detection cross section models.
In Section~\ref{sec:lifetime}
we have discussed the dependence of the results of the analysis of the Gallium data on the
assumed value of the half life of ${}^{71}\text{Ge}$,
which determines the cross sections of the transitions
from the ground state of
${}^{71}\text{Ga}$ to the ground state of ${}^{71}\text{Ge}$.
The standard value of the ${}^{71}\text{Ge}$ half life was
measured in 1985,
but previous measurements gave different values.
In Section~\ref{sec:dependence} we have shown that the Gallium Anomaly
can be reduced or solved with a value of the ${}^{71}\text{Ge}$ half life
which is larger than the standard one.
In Section~\ref{sec:oscillations}
we have considered the short-baseline neutrino oscillation interpretation of the Gallium Anomaly.
We have shown that a value of the ${}^{71}\text{Ge}$ half life
which is larger than the standard one can reduce the tension between the Gallium data and the results
of other experiments
(short-baseline reactor neutrino experiments, solar neutrinos,
and $\beta$-decay data)~\cite{Berryman:2021yan,Giunti:2022btk}.

We conclude by emphasizing that new measurements of the ${}^{71}\text{Ge}$ half life with modern technique and apparatus are needed for a better assessment of the Gallium Anomaly.

\begin{acknowledgments}
We would like to thank David Lhuillier for stimulating discussions.
C.G. and C.A.T. are supported by the research grant ``The Dark Universe: A Synergic Multimessenger Approach'' number 2017X7X85K under the program ``PRIN 2017'' funded by the Italian Ministero dell'Istruzione, Universit\`a e della Ricerca (MIUR). C.A.T. also acknowledges support from {\sl Departments of Excellence} grant awarded by MIUR and the research grant {\sl TAsP (Theoretical Astroparticle Physics)} funded by Istituto Nazionale di Fisica Nucleare (INFN).
The work of Y.F.Li and Z.Xin was supported by National Natural Science Foundation of China under Grant Nos.~12075255 and 11835013, by the Key Research Program of the Chinese Academy of Sciences under Grant No.~XDPB15.
\end{acknowledgments}

%%%%%%%%%%%%%%%%%%%%%%%%%%%%%%%%%%%%%%%%%%%%%%%%%%%%%%%%

%\bibliographystyle{apsrev4-1}
%\bibliography{main}

\begin{thebibliography}{43}%
\makeatletter
\providecommand \@ifxundefined [1]{%
 \@ifx{#1\undefined}
}%
\providecommand \@ifnum [1]{%
 \ifnum #1\expandafter \@firstoftwo
 \else \expandafter \@secondoftwo
 \fi
}%
\providecommand \@ifx [1]{%
 \ifx #1\expandafter \@firstoftwo
 \else \expandafter \@secondoftwo
 \fi
}%
\providecommand \natexlab [1]{#1}%
\providecommand \enquote  [1]{``#1''}%
\providecommand \bibnamefont  [1]{#1}%
\providecommand \bibfnamefont [1]{#1}%
\providecommand \citenamefont [1]{#1}%
\providecommand \href@noop [0]{\@secondoftwo}%
\providecommand \href [0]{\begingroup \@sanitize@url \@href}%
\providecommand \@href[1]{\@@startlink{#1}\@@href}%
\providecommand \@@href[1]{\endgroup#1\@@endlink}%
\providecommand \@sanitize@url [0]{\catcode `\\12\catcode `\$12\catcode
  `\&12\catcode `\#12\catcode `\^12\catcode `\_12\catcode `\%12\relax}%
\providecommand \@@startlink[1]{}%
\providecommand \@@endlink[0]{}%
\providecommand \url  [0]{\begingroup\@sanitize@url \@url }%
\providecommand \@url [1]{\endgroup\@href {#1}{\urlprefix }}%
\providecommand \urlprefix  [0]{URL }%
\providecommand \Eprint [0]{\href }%
\providecommand \doibase [0]{http://dx.doi.org/}%
\providecommand \selectlanguage [0]{\@gobble}%
\providecommand \bibinfo  [0]{\@secondoftwo}%
\providecommand \bibfield  [0]{\@secondoftwo}%
\providecommand \translation [1]{[#1]}%
\providecommand \BibitemOpen [0]{}%
\providecommand \bibitemStop [0]{}%
\providecommand \bibitemNoStop [0]{.\EOS\space}%
\providecommand \EOS [0]{\spacefactor3000\relax}%
\providecommand \BibitemShut  [1]{\csname bibitem#1\endcsname}%
\let\auto@bib@innerbib\@empty
%</preamble>
\bibitem [{\citenamefont {Abdurashitov}\ \emph {et~al.}(2006)\citenamefont
  {Abdurashitov} \emph {et~al.}}]{Abdurashitov:2005tb}%
  \BibitemOpen
  \bibfield  {author} {\bibinfo {author} {\bibfnamefont {J.~N.}\ \bibnamefont
  {Abdurashitov}} \emph {et~al.} (\bibinfo {collaboration} {SAGE}),\
  }\href@noop {} {\bibfield  {journal} {\bibinfo  {journal} {Phys. Rev.}\
  }\textbf {\bibinfo {volume} {C73}},\ \bibinfo {pages} {045805} (\bibinfo
  {year} {2006})},\ \Eprint {http://arxiv.org/abs/nucl-ex/0512041}
  {nucl-ex/0512041} \BibitemShut {NoStop}%
\bibitem [{\citenamefont {Laveder}(2007)}]{Laveder:2007zz}%
  \BibitemOpen
  \bibfield  {author} {\bibinfo {author} {\bibfnamefont {M.}~\bibnamefont
  {Laveder}},\ }\href {\doibase 10.1016/j.nuclphysbps.2007.02.037} {\bibfield
  {journal} {\bibinfo  {journal} {Nucl. Phys. Proc. Suppl.}\ }\textbf {\bibinfo
  {volume} {168}},\ \bibinfo {pages} {344} (\bibinfo {year}
  {2007})}\BibitemShut {NoStop}%
\bibitem [{\citenamefont {Giunti}\ and\ \citenamefont
  {Laveder}(2007)}]{Giunti:2006bj}%
  \BibitemOpen
  \bibfield  {author} {\bibinfo {author} {\bibfnamefont {C.}~\bibnamefont
  {Giunti}}\ and\ \bibinfo {author} {\bibfnamefont {M.}~\bibnamefont
  {Laveder}},\ }\href@noop {} {\bibfield  {journal} {\bibinfo  {journal} {Mod.
  Phys. Lett.}\ }\textbf {\bibinfo {volume} {A22}},\ \bibinfo {pages} {2499}
  (\bibinfo {year} {2007})},\ \Eprint {http://arxiv.org/abs/hep-ph/0610352}
  {hep-ph/0610352} \BibitemShut {NoStop}%
\bibitem [{\citenamefont {Anselmann}\ \emph {et~al.}(1995)\citenamefont
  {Anselmann} \emph {et~al.}}]{GALLEX:1994rym}%
  \BibitemOpen
  \bibfield  {author} {\bibinfo {author} {\bibfnamefont {P.}~\bibnamefont
  {Anselmann}} \emph {et~al.} (\bibinfo {collaboration} {GALLEX}),\ }\href@noop
  {} {\bibfield  {journal} {\bibinfo  {journal} {Phys. Lett.}\ }\textbf
  {\bibinfo {volume} {B342}},\ \bibinfo {pages} {440} (\bibinfo {year}
  {1995})}\BibitemShut {NoStop}%
\bibitem [{\citenamefont {Hampel}\ \emph {et~al.}(1998)\citenamefont {Hampel}
  \emph {et~al.}}]{GALLEX:1997lja}%
  \BibitemOpen
  \bibfield  {author} {\bibinfo {author} {\bibfnamefont {W.}~\bibnamefont
  {Hampel}} \emph {et~al.} (\bibinfo {collaboration} {GALLEX}),\ }\href@noop {}
  {\bibfield  {journal} {\bibinfo  {journal} {Phys. Lett.}\ }\textbf {\bibinfo
  {volume} {B420}},\ \bibinfo {pages} {114} (\bibinfo {year}
  {1998})}\BibitemShut {NoStop}%
\bibitem [{\citenamefont {Kaether}\ \emph {et~al.}(2010)\citenamefont
  {Kaether}, \citenamefont {Hampel}, \citenamefont {Heusser}, \citenamefont
  {Kiko},\ and\ \citenamefont {Kirsten}}]{Kaether:2010ag}%
  \BibitemOpen
  \bibfield  {author} {\bibinfo {author} {\bibfnamefont {F.}~\bibnamefont
  {Kaether}}, \bibinfo {author} {\bibfnamefont {W.}~\bibnamefont {Hampel}},
  \bibinfo {author} {\bibfnamefont {G.}~\bibnamefont {Heusser}}, \bibinfo
  {author} {\bibfnamefont {J.}~\bibnamefont {Kiko}}, \ and\ \bibinfo {author}
  {\bibfnamefont {T.}~\bibnamefont {Kirsten}},\ }\href@noop {} {\bibfield
  {journal} {\bibinfo  {journal} {Phys. Lett.}\ }\textbf {\bibinfo {volume}
  {B685}},\ \bibinfo {pages} {47} (\bibinfo {year} {2010})},\ \Eprint
  {http://arxiv.org/abs/arXiv:1001.2731} {arXiv:1001.2731 [hep-ex]}
  \BibitemShut {NoStop}%
\bibitem [{\citenamefont {Abdurashitov}\ \emph {et~al.}(1996)\citenamefont
  {Abdurashitov} \emph {et~al.}}]{Abdurashitov:1996dp}%
  \BibitemOpen
  \bibfield  {author} {\bibinfo {author} {\bibfnamefont {J.~N.}\ \bibnamefont
  {Abdurashitov}} \emph {et~al.} (\bibinfo {collaboration} {SAGE}),\
  }\href@noop {} {\bibfield  {journal} {\bibinfo  {journal} {Phys. Rev. Lett.}\
  }\textbf {\bibinfo {volume} {77}},\ \bibinfo {pages} {4708} (\bibinfo {year}
  {1996})}\BibitemShut {NoStop}%
\bibitem [{\citenamefont {Abdurashitov}\ \emph {et~al.}(1999)\citenamefont
  {Abdurashitov} \emph {et~al.}}]{SAGE:1998fvr}%
  \BibitemOpen
  \bibfield  {author} {\bibinfo {author} {\bibfnamefont {J.~N.}\ \bibnamefont
  {Abdurashitov}} \emph {et~al.} (\bibinfo {collaboration} {SAGE}),\
  }\href@noop {} {\bibfield  {journal} {\bibinfo  {journal} {Phys. Rev.}\
  }\textbf {\bibinfo {volume} {C59}},\ \bibinfo {pages} {2246} (\bibinfo {year}
  {1999})},\ \Eprint {http://arxiv.org/abs/hep-ph/9803418} {hep-ph/9803418}
  \BibitemShut {NoStop}%
\bibitem [{\citenamefont {Abdurashitov}\ \emph {et~al.}(2009)\citenamefont
  {Abdurashitov} \emph {et~al.}}]{SAGE:2009eeu}%
  \BibitemOpen
  \bibfield  {author} {\bibinfo {author} {\bibfnamefont {J.~N.}\ \bibnamefont
  {Abdurashitov}} \emph {et~al.} (\bibinfo {collaboration} {SAGE}),\
  }\href@noop {} {\bibfield  {journal} {\bibinfo  {journal} {Phys. Rev.}\
  }\textbf {\bibinfo {volume} {C80}},\ \bibinfo {pages} {015807} (\bibinfo
  {year} {2009})},\ \Eprint {http://arxiv.org/abs/arXiv:0901.2200}
  {arXiv:0901.2200 [nucl-ex]} \BibitemShut {NoStop}%
\bibitem [{\citenamefont {Bahcall}(1997)}]{Bahcall:1997eg}%
  \BibitemOpen
  \bibfield  {author} {\bibinfo {author} {\bibfnamefont {J.~N.}\ \bibnamefont
  {Bahcall}},\ }\href@noop {} {\bibfield  {journal} {\bibinfo  {journal} {Phys.
  Rev.}\ }\textbf {\bibinfo {volume} {C56}},\ \bibinfo {pages} {3391} (\bibinfo
  {year} {1997})},\ \Eprint {http://arxiv.org/abs/hep-ph/9710491}
  {hep-ph/9710491} \BibitemShut {NoStop}%
\bibitem [{\citenamefont {Gariazzo}\ \emph {et~al.}(2016)\citenamefont
  {Gariazzo}, \citenamefont {Giunti}, \citenamefont {Laveder}, \citenamefont
  {Li},\ and\ \citenamefont {Zavanin}}]{Gariazzo:2015rra}%
  \BibitemOpen
  \bibfield  {author} {\bibinfo {author} {\bibfnamefont {S.}~\bibnamefont
  {Gariazzo}}, \bibinfo {author} {\bibfnamefont {C.}~\bibnamefont {Giunti}},
  \bibinfo {author} {\bibfnamefont {M.}~\bibnamefont {Laveder}}, \bibinfo
  {author} {\bibfnamefont {Y.~F.}\ \bibnamefont {Li}}, \ and\ \bibinfo {author}
  {\bibfnamefont {E.}~\bibnamefont {Zavanin}},\ }\href@noop {} {\bibfield
  {journal} {\bibinfo  {journal} {J. Phys.}\ }\textbf {\bibinfo {volume}
  {G43}},\ \bibinfo {pages} {033001} (\bibinfo {year} {2016})},\ \Eprint
  {http://arxiv.org/abs/arXiv:1507.08204} {arXiv:1507.08204 [hep-ph]}
  \BibitemShut {NoStop}%
\bibitem [{\citenamefont {Gonzalez-Garcia}\ \emph {et~al.}(2016)\citenamefont
  {Gonzalez-Garcia}, \citenamefont {Maltoni},\ and\ \citenamefont
  {Schwetz}}]{Gonzalez-Garcia:2015qrr}%
  \BibitemOpen
  \bibfield  {author} {\bibinfo {author} {\bibfnamefont {M.}~\bibnamefont
  {Gonzalez-Garcia}}, \bibinfo {author} {\bibfnamefont {M.}~\bibnamefont
  {Maltoni}}, \ and\ \bibinfo {author} {\bibfnamefont {T.}~\bibnamefont
  {Schwetz}},\ }\href@noop {} {\bibfield  {journal} {\bibinfo  {journal} {Nucl.
  Phys.}\ }\textbf {\bibinfo {volume} {B908}},\ \bibinfo {pages} {199}
  (\bibinfo {year} {2016})},\ \Eprint {http://arxiv.org/abs/arXiv:1512.06856}
  {arXiv:1512.06856 [hep-ph]} \BibitemShut {NoStop}%
\bibitem [{\citenamefont {Giunti}\ and\ \citenamefont
  {Lasserre}(2019)}]{Giunti:2019aiy}%
  \BibitemOpen
  \bibfield  {author} {\bibinfo {author} {\bibfnamefont {C.}~\bibnamefont
  {Giunti}}\ and\ \bibinfo {author} {\bibfnamefont {T.}~\bibnamefont
  {Lasserre}},\ }\href {\doibase 10.1146/annurev-nucl-101918-023755} {\bibfield
   {journal} {\bibinfo  {journal} {Ann. Rev. Nucl. Part. Sci.}\ }\textbf
  {\bibinfo {volume} {69}},\ \bibinfo {pages} {163} (\bibinfo {year} {2019})},\
  \Eprint {http://arxiv.org/abs/arXiv:1901.08330} {arXiv:1901.08330 [hep-ph]}
  \BibitemShut {NoStop}%
\bibitem [{\citenamefont {Diaz}\ \emph {et~al.}(2020)\citenamefont {Diaz},
  \citenamefont {Arguelles}, \citenamefont {Collin}, \citenamefont {Conrad},\
  and\ \citenamefont {Shaevitz}}]{Diaz:2019fwt}%
  \BibitemOpen
  \bibfield  {author} {\bibinfo {author} {\bibfnamefont {A.}~\bibnamefont
  {Diaz}}, \bibinfo {author} {\bibfnamefont {C.}~\bibnamefont {Arguelles}},
  \bibinfo {author} {\bibfnamefont {G.}~\bibnamefont {Collin}}, \bibinfo
  {author} {\bibfnamefont {J.}~\bibnamefont {Conrad}}, \ and\ \bibinfo {author}
  {\bibfnamefont {M.}~\bibnamefont {Shaevitz}},\ }\href@noop {} {\bibfield
  {journal} {\bibinfo  {journal} {Phys.Rept.}\ }\textbf {\bibinfo {volume}
  {884}},\ \bibinfo {pages} {1} (\bibinfo {year} {2020})},\ \Eprint
  {http://arxiv.org/abs/arXiv:1906.00045} {arXiv:1906.00045 [hep-ex]}
  \BibitemShut {NoStop}%
\bibitem [{\citenamefont {Boser}\ \emph {et~al.}(2020)\citenamefont {Boser},
  \citenamefont {Buck}, \citenamefont {Giunti}, \citenamefont {Lesgourgues},
  \citenamefont {Ludhova}, \citenamefont {Mertens}, \citenamefont {Schukraft},\
  and\ \citenamefont {Wurm}}]{Boser:2019rta}%
  \BibitemOpen
  \bibfield  {author} {\bibinfo {author} {\bibfnamefont {S.}~\bibnamefont
  {Boser}}, \bibinfo {author} {\bibfnamefont {C.}~\bibnamefont {Buck}},
  \bibinfo {author} {\bibfnamefont {C.}~\bibnamefont {Giunti}}, \bibinfo
  {author} {\bibfnamefont {J.}~\bibnamefont {Lesgourgues}}, \bibinfo {author}
  {\bibfnamefont {L.}~\bibnamefont {Ludhova}}, \bibinfo {author} {\bibfnamefont
  {S.}~\bibnamefont {Mertens}}, \bibinfo {author} {\bibfnamefont
  {A.}~\bibnamefont {Schukraft}}, \ and\ \bibinfo {author} {\bibfnamefont
  {M.}~\bibnamefont {Wurm}},\ }\href@noop {} {\bibfield  {journal} {\bibinfo
  {journal} {Prog.Part.Nucl.Phys.}\ }\textbf {\bibinfo {volume} {111}},\
  \bibinfo {pages} {103736} (\bibinfo {year} {2020})},\ \Eprint
  {http://arxiv.org/abs/arXiv:1906.01739} {arXiv:1906.01739 [hep-ex]}
  \BibitemShut {NoStop}%
\bibitem [{\citenamefont {Dasgupta}\ and\ \citenamefont
  {Kopp}(2021)}]{Dasgupta:2021ies}%
  \BibitemOpen
  \bibfield  {author} {\bibinfo {author} {\bibfnamefont {B.}~\bibnamefont
  {Dasgupta}}\ and\ \bibinfo {author} {\bibfnamefont {J.}~\bibnamefont
  {Kopp}},\ }\href@noop {} {\bibfield  {journal} {\bibinfo  {journal}
  {Phys.Rept.}\ }\textbf {\bibinfo {volume} {928}},\ \bibinfo {pages} {63}
  (\bibinfo {year} {2021})},\ \Eprint {http://arxiv.org/abs/arXiv:2106.05913}
  {arXiv:2106.05913 [hep-ph]} \BibitemShut {NoStop}%
\bibitem [{\citenamefont {Barinov}\ \emph
  {et~al.}(2022{\natexlab{a}})\citenamefont {Barinov} \emph
  {et~al.}}]{Barinov:2021asz}%
  \BibitemOpen
  \bibfield  {author} {\bibinfo {author} {\bibfnamefont {V.}~\bibnamefont
  {Barinov}} \emph {et~al.} (\bibinfo {collaboration} {BEST}),\ }\href@noop {}
  {\bibfield  {journal} {\bibinfo  {journal} {Phys.Rev.Lett.}\ }\textbf
  {\bibinfo {volume} {128}},\ \bibinfo {pages} {232501} (\bibinfo {year}
  {2022}{\natexlab{a}})},\ \Eprint {http://arxiv.org/abs/arXiv:2109.11482}
  {arXiv:2109.11482 [nucl-ex]} \BibitemShut {NoStop}%
\bibitem [{\citenamefont {Barinov}\ \emph
  {et~al.}(2022{\natexlab{b}})\citenamefont {Barinov} \emph
  {et~al.}}]{Barinov:2022wfh}%
  \BibitemOpen
  \bibfield  {author} {\bibinfo {author} {\bibfnamefont {V.}~\bibnamefont
  {Barinov}} \emph {et~al.} (\bibinfo {collaboration} {BEST}),\ }\href@noop {}
  {\bibfield  {journal} {\bibinfo  {journal} {Phys.Rev.C}\ }\textbf {\bibinfo
  {volume} {105}},\ \bibinfo {pages} {065502} (\bibinfo {year}
  {2022}{\natexlab{b}})},\ \Eprint {http://arxiv.org/abs/arXiv:2201.07364}
  {arXiv:2201.07364 [nucl-ex]} \BibitemShut {NoStop}%
\bibitem [{\citenamefont {Giunti}\ \emph
  {et~al.}(2022{\natexlab{a}})\citenamefont {Giunti}, \citenamefont {Li},
  \citenamefont {Ternes}, \citenamefont {Tyagi},\ and\ \citenamefont
  {Xin}}]{Giunti:2022btk}%
  \BibitemOpen
  \bibfield  {author} {\bibinfo {author} {\bibfnamefont {C.}~\bibnamefont
  {Giunti}}, \bibinfo {author} {\bibfnamefont {Y.~F.}\ \bibnamefont {Li}},
  \bibinfo {author} {\bibfnamefont {C.~A.}\ \bibnamefont {Ternes}}, \bibinfo
  {author} {\bibfnamefont {O.}~\bibnamefont {Tyagi}}, \ and\ \bibinfo {author}
  {\bibfnamefont {Z.}~\bibnamefont {Xin}},\ }\href@noop {} {\bibfield
  {journal} {\bibinfo  {journal} {JHEP}\ }\textbf {\bibinfo {volume} {10}},\
  \bibinfo {pages} {164} (\bibinfo {year} {2022}{\natexlab{a}})},\ \Eprint
  {http://arxiv.org/abs/arXiv:2209.00916} {arXiv:2209.00916 [hep-ph]}
  \BibitemShut {NoStop}%
\bibitem [{\citenamefont {Workman}(2022)}]{ParticleDataGroup:2022pth}%
  \BibitemOpen
  \bibfield  {author} {\bibinfo {author} {\bibfnamefont {R.~L.}\ \bibnamefont
  {Workman}} (\bibinfo {collaboration} {Particle Data Group}),\ }\href@noop {}
  {\bibfield  {journal} {\bibinfo  {journal} {PTEP}\ }\textbf {\bibinfo
  {volume} {2022}},\ \bibinfo {pages} {083C01} (\bibinfo {year}
  {2022})}\BibitemShut {NoStop}%
\bibitem [{\citenamefont {de~Salas}\ \emph {et~al.}(2020)\citenamefont
  {de~Salas}, \citenamefont {Forero}, \citenamefont {Gariazzo}, \citenamefont
  {Martinez-Mirave}, \citenamefont {Mena}, \citenamefont {Ternes},
  \citenamefont {Tortola},\ and\ \citenamefont {Valle}}]{deSalas:2020pgw}%
  \BibitemOpen
  \bibfield  {author} {\bibinfo {author} {\bibfnamefont {P.~F.}\ \bibnamefont
  {de~Salas}}, \bibinfo {author} {\bibfnamefont {D.~V.}\ \bibnamefont
  {Forero}}, \bibinfo {author} {\bibfnamefont {S.}~\bibnamefont {Gariazzo}},
  \bibinfo {author} {\bibfnamefont {P.}~\bibnamefont {Martinez-Mirave}},
  \bibinfo {author} {\bibfnamefont {O.}~\bibnamefont {Mena}}, \bibinfo {author}
  {\bibfnamefont {C.~A.}\ \bibnamefont {Ternes}}, \bibinfo {author}
  {\bibfnamefont {M.}~\bibnamefont {Tortola}}, \ and\ \bibinfo {author}
  {\bibfnamefont {J.~W.~F.}\ \bibnamefont {Valle}},\ }\href@noop {} {\bibfield
  {journal} {\bibinfo  {journal} {JHEP}\ }\textbf {\bibinfo {volume} {2021}},\
  \bibinfo {pages} {071} (\bibinfo {year} {2020})},\ \Eprint
  {http://arxiv.org/abs/arXiv:2006.11237} {arXiv:2006.11237 [hep-ph]}
  \BibitemShut {NoStop}%
\bibitem [{\citenamefont {Capozzi}\ \emph {et~al.}(2021)\citenamefont
  {Capozzi}, \citenamefont {Di~Valentino}, \citenamefont {Lisi}, \citenamefont
  {Marrone}, \citenamefont {Melchiorri},\ and\ \citenamefont
  {Palazzo}}]{Capozzi:2021fjo}%
  \BibitemOpen
  \bibfield  {author} {\bibinfo {author} {\bibfnamefont {F.}~\bibnamefont
  {Capozzi}}, \bibinfo {author} {\bibfnamefont {E.}~\bibnamefont
  {Di~Valentino}}, \bibinfo {author} {\bibfnamefont {E.}~\bibnamefont {Lisi}},
  \bibinfo {author} {\bibfnamefont {A.}~\bibnamefont {Marrone}}, \bibinfo
  {author} {\bibfnamefont {A.}~\bibnamefont {Melchiorri}}, \ and\ \bibinfo
  {author} {\bibfnamefont {A.}~\bibnamefont {Palazzo}},\ }\href {\doibase
  10.1103/PhysRevD.104.083031} {\bibfield  {journal} {\bibinfo  {journal}
  {Phys. Rev. D}\ }\textbf {\bibinfo {volume} {104}},\ \bibinfo {pages}
  {083031} (\bibinfo {year} {2021})},\ \Eprint
  {http://arxiv.org/abs/2107.00532} {arXiv:2107.00532 [hep-ph]} \BibitemShut
  {NoStop}%
\bibitem [{\citenamefont {Esteban}\ \emph {et~al.}(2020)\citenamefont
  {Esteban}, \citenamefont {Gonzalez-Garcia}, \citenamefont {Maltoni},
  \citenamefont {Schwetz},\ and\ \citenamefont {Zhou}}]{Esteban:2020cvm}%
  \BibitemOpen
  \bibfield  {author} {\bibinfo {author} {\bibfnamefont {I.}~\bibnamefont
  {Esteban}}, \bibinfo {author} {\bibfnamefont {M.~C.}\ \bibnamefont
  {Gonzalez-Garcia}}, \bibinfo {author} {\bibfnamefont {M.}~\bibnamefont
  {Maltoni}}, \bibinfo {author} {\bibfnamefont {T.}~\bibnamefont {Schwetz}}, \
  and\ \bibinfo {author} {\bibfnamefont {A.}~\bibnamefont {Zhou}},\ }\href
  {\doibase 10.1007/JHEP09(2020)178} {\bibfield  {journal} {\bibinfo  {journal}
  {JHEP}\ }\textbf {\bibinfo {volume} {09}},\ \bibinfo {pages} {178} (\bibinfo
  {year} {2020})},\ \Eprint {http://arxiv.org/abs/2007.14792} {arXiv:2007.14792
  [hep-ph]} \BibitemShut {NoStop}%
\bibitem [{\citenamefont {Berryman}\ \emph {et~al.}(2022)\citenamefont
  {Berryman}, \citenamefont {Coloma}, \citenamefont {Huber}, \citenamefont
  {Schwetz},\ and\ \citenamefont {Zhou}}]{Berryman:2021yan}%
  \BibitemOpen
  \bibfield  {author} {\bibinfo {author} {\bibfnamefont {J.~M.}\ \bibnamefont
  {Berryman}}, \bibinfo {author} {\bibfnamefont {P.}~\bibnamefont {Coloma}},
  \bibinfo {author} {\bibfnamefont {P.}~\bibnamefont {Huber}}, \bibinfo
  {author} {\bibfnamefont {T.}~\bibnamefont {Schwetz}}, \ and\ \bibinfo
  {author} {\bibfnamefont {A.}~\bibnamefont {Zhou}},\ }\href@noop {} {\bibfield
   {journal} {\bibinfo  {journal} {JHEP}\ }\textbf {\bibinfo {volume} {02}},\
  \bibinfo {pages} {055} (\bibinfo {year} {2022})},\ \Eprint
  {http://arxiv.org/abs/arXiv:2111.12530} {arXiv:2111.12530 [hep-ph]}
  \BibitemShut {NoStop}%
\bibitem [{\citenamefont {Giunti}\ \emph
  {et~al.}(2022{\natexlab{b}})\citenamefont {Giunti}, \citenamefont {Li},
  \citenamefont {Ternes},\ and\ \citenamefont {Xin}}]{Giunti:2021kab}%
  \BibitemOpen
  \bibfield  {author} {\bibinfo {author} {\bibfnamefont {C.}~\bibnamefont
  {Giunti}}, \bibinfo {author} {\bibfnamefont {Y.}~\bibnamefont {Li}}, \bibinfo
  {author} {\bibfnamefont {C.}~\bibnamefont {Ternes}}, \ and\ \bibinfo {author}
  {\bibfnamefont {Z.}~\bibnamefont {Xin}},\ }\href@noop {} {\bibfield
  {journal} {\bibinfo  {journal} {Phys.Lett.B}\ }\textbf {\bibinfo {volume}
  {829}},\ \bibinfo {pages} {137054} (\bibinfo {year} {2022}{\natexlab{b}})},\
  \Eprint {http://arxiv.org/abs/arXiv:2110.06820} {arXiv:2110.06820 [hep-ph]}
  \BibitemShut {NoStop}%
\bibitem [{\citenamefont {Serebrov}\ \emph {et~al.}(2021)\citenamefont
  {Serebrov} \emph {et~al.}}]{Serebrov:2020kmd}%
  \BibitemOpen
  \bibfield  {author} {\bibinfo {author} {\bibfnamefont {A.~P.}\ \bibnamefont
  {Serebrov}} \emph {et~al.},\ }\href {\doibase 10.1103/PhysRevD.104.032003}
  {\bibfield  {journal} {\bibinfo  {journal} {Phys. Rev. D}\ }\textbf {\bibinfo
  {volume} {104}},\ \bibinfo {pages} {032003} (\bibinfo {year} {2021})},\
  \Eprint {http://arxiv.org/abs/2005.05301} {arXiv:2005.05301 [hep-ex]}
  \BibitemShut {NoStop}%
\bibitem [{\citenamefont {Barinov}\ and\ \citenamefont
  {Gorbunov}(2022)}]{Barinov:2021mjj}%
  \BibitemOpen
  \bibfield  {author} {\bibinfo {author} {\bibfnamefont {V.}~\bibnamefont
  {Barinov}}\ and\ \bibinfo {author} {\bibfnamefont {D.}~\bibnamefont
  {Gorbunov}},\ }\href {\doibase 10.1103/PhysRevD.105.L051703} {\bibfield
  {journal} {\bibinfo  {journal} {Phys. Rev. D}\ }\textbf {\bibinfo {volume}
  {105}},\ \bibinfo {pages} {L051703} (\bibinfo {year} {2022})},\ \Eprint
  {http://arxiv.org/abs/2109.14654} {arXiv:2109.14654 [hep-ph]} \BibitemShut
  {NoStop}%
\bibitem [{\citenamefont {Danilov}\ and\ \citenamefont
  {Skrobova}(2020)}]{Danilov:2020rax}%
  \BibitemOpen
  \bibfield  {author} {\bibinfo {author} {\bibfnamefont {M.~V.}\ \bibnamefont
  {Danilov}}\ and\ \bibinfo {author} {\bibfnamefont {N.~A.}\ \bibnamefont
  {Skrobova}},\ }\href {\doibase 10.1134/S0021364020190066} {\bibfield
  {journal} {\bibinfo  {journal} {JETP Lett.}\ }\textbf {\bibinfo {volume}
  {112}},\ \bibinfo {pages} {452} (\bibinfo {year} {2020})}\BibitemShut
  {NoStop}%
\bibitem [{\citenamefont {Giunti}\ \emph {et~al.}(2021)\citenamefont {Giunti},
  \citenamefont {Li}, \citenamefont {Ternes},\ and\ \citenamefont
  {Zhang}}]{Giunti:2021iti}%
  \BibitemOpen
  \bibfield  {author} {\bibinfo {author} {\bibfnamefont {C.}~\bibnamefont
  {Giunti}}, \bibinfo {author} {\bibfnamefont {Y.~F.}\ \bibnamefont {Li}},
  \bibinfo {author} {\bibfnamefont {C.~A.}\ \bibnamefont {Ternes}}, \ and\
  \bibinfo {author} {\bibfnamefont {Y.~Y.}\ \bibnamefont {Zhang}},\ }\href
  {\doibase 10.1016/j.physletb.2021.136214} {\bibfield  {journal} {\bibinfo
  {journal} {Phys. Lett. B}\ }\textbf {\bibinfo {volume} {816}},\ \bibinfo
  {pages} {136214} (\bibinfo {year} {2021})},\ \Eprint
  {http://arxiv.org/abs/2101.06785} {arXiv:2101.06785 [hep-ph]} \BibitemShut
  {NoStop}%
\bibitem [{\citenamefont {Haxton}(1998)}]{Haxton:1998uc}%
  \BibitemOpen
  \bibfield  {author} {\bibinfo {author} {\bibfnamefont {W.~C.}\ \bibnamefont
  {Haxton}},\ }\href {\doibase 10.1016/S0370-2693(98)00581-4} {\bibfield
  {journal} {\bibinfo  {journal} {Phys. Lett.}\ }\textbf {\bibinfo {volume}
  {B431}},\ \bibinfo {pages} {110} (\bibinfo {year} {1998})},\ \Eprint
  {http://arxiv.org/abs/nucl-th/9804011} {nucl-th/9804011} \BibitemShut
  {NoStop}%
\bibitem [{\citenamefont {Frekers}\ \emph {et~al.}(2015)\citenamefont {Frekers}
  \emph {et~al.}}]{Frekers:2015wga}%
  \BibitemOpen
  \bibfield  {author} {\bibinfo {author} {\bibfnamefont {D.}~\bibnamefont
  {Frekers}} \emph {et~al.},\ }\href {\doibase 10.1103/PhysRevC.91.034608}
  {\bibfield  {journal} {\bibinfo  {journal} {Phys. Rev.}\ }\textbf {\bibinfo
  {volume} {C91}},\ \bibinfo {pages} {034608} (\bibinfo {year}
  {2015})}\BibitemShut {NoStop}%
\bibitem [{\citenamefont {Kostensalo}\ \emph {et~al.}(2019)\citenamefont
  {Kostensalo}, \citenamefont {Suhonen}, \citenamefont {Giunti},\ and\
  \citenamefont {Srivastava}}]{Kostensalo:2019vmv}%
  \BibitemOpen
  \bibfield  {author} {\bibinfo {author} {\bibfnamefont {J.}~\bibnamefont
  {Kostensalo}}, \bibinfo {author} {\bibfnamefont {J.}~\bibnamefont {Suhonen}},
  \bibinfo {author} {\bibfnamefont {C.}~\bibnamefont {Giunti}}, \ and\ \bibinfo
  {author} {\bibfnamefont {P.~C.}\ \bibnamefont {Srivastava}},\ }\href@noop {}
  {\bibfield  {journal} {\bibinfo  {journal} {Phys.Lett.}\ }\textbf {\bibinfo
  {volume} {B795}},\ \bibinfo {pages} {542} (\bibinfo {year} {2019})},\ \Eprint
  {http://arxiv.org/abs/arXiv:1906.10980} {arXiv:1906.10980 [nucl-th]}
  \BibitemShut {NoStop}%
\bibitem [{\citenamefont {Semenov}(2020)}]{Semenov:2020xea}%
  \BibitemOpen
  \bibfield  {author} {\bibinfo {author} {\bibfnamefont {S.~V.}\ \bibnamefont
  {Semenov}},\ }\href {\doibase 10.1134/S1063778820100221} {\bibfield
  {journal} {\bibinfo  {journal} {Phys. Atom. Nucl.}\ }\textbf {\bibinfo
  {volume} {83}},\ \bibinfo {pages} {1549} (\bibinfo {year}
  {2020})}\BibitemShut {NoStop}%
\bibitem [{\citenamefont {Krofcheck}\ \emph {et~al.}(1985)\citenamefont
  {Krofcheck} \emph {et~al.}}]{Krofcheck:1985fg}%
  \BibitemOpen
  \bibfield  {author} {\bibinfo {author} {\bibfnamefont {D.}~\bibnamefont
  {Krofcheck}} \emph {et~al.},\ }\href {\doibase 10.1103/PhysRevLett.55.1051}
  {\bibfield  {journal} {\bibinfo  {journal} {Phys. Rev. Lett.}\ }\textbf
  {\bibinfo {volume} {55}},\ \bibinfo {pages} {1051} (\bibinfo {year}
  {1985})}\BibitemShut {NoStop}%
\bibitem [{\citenamefont {{Krofcheck}}(1987)}]{Krofcheck-PhD-1987}%
  \BibitemOpen
  \bibfield  {author} {\bibinfo {author} {\bibfnamefont {D.}~\bibnamefont
  {{Krofcheck}}},\ }\href@noop {} {\  (\bibinfo {year} {1987})},\ \bibinfo
  {note} {phD Thesis, Ohio State University}\BibitemShut {NoStop}%
\bibitem [{\citenamefont {Frekers}\ \emph {et~al.}(2011)\citenamefont
  {Frekers}, \citenamefont {Ejiri}, \citenamefont {Akimune}, \citenamefont
  {Adachi}, \citenamefont {Bilgier} \emph {et~al.}}]{Frekers:2011zz}%
  \BibitemOpen
  \bibfield  {author} {\bibinfo {author} {\bibfnamefont {D.}~\bibnamefont
  {Frekers}}, \bibinfo {author} {\bibfnamefont {H.}~\bibnamefont {Ejiri}},
  \bibinfo {author} {\bibfnamefont {H.}~\bibnamefont {Akimune}}, \bibinfo
  {author} {\bibfnamefont {T.}~\bibnamefont {Adachi}}, \bibinfo {author}
  {\bibfnamefont {B.}~\bibnamefont {Bilgier}},  \emph {et~al.},\ }\href@noop {}
  {\bibfield  {journal} {\bibinfo  {journal} {Phys. Lett.}\ }\textbf {\bibinfo
  {volume} {B706}},\ \bibinfo {pages} {134} (\bibinfo {year}
  {2011})}\BibitemShut {NoStop}%
\bibitem [{\citenamefont {Bahcall}(1978)}]{Bahcall:1978fa}%
  \BibitemOpen
  \bibfield  {author} {\bibinfo {author} {\bibfnamefont {J.~N.}\ \bibnamefont
  {Bahcall}},\ }\href {\doibase 10.1103/RevModPhys.50.881} {\bibfield
  {journal} {\bibinfo  {journal} {Rev. Mod. Phys.}\ }\textbf {\bibinfo {volume}
  {50}},\ \bibinfo {pages} {881} (\bibinfo {year} {1978})}\BibitemShut
  {NoStop}%
\bibitem [{\citenamefont {Alanssari}\ \emph {et~al.}(2016)\citenamefont
  {Alanssari} \emph {et~al.}}]{Alanssari:2016itw}%
  \BibitemOpen
  \bibfield  {author} {\bibinfo {author} {\bibfnamefont {M.}~\bibnamefont
  {Alanssari}} \emph {et~al.},\ }\href {\doibase 10.1016/j.ijms.2016.05.019}
  {\bibfield  {journal} {\bibinfo  {journal} {Int. J. Mass Spectrometry}\
  }\textbf {\bibinfo {volume} {406}},\ \bibinfo {pages} {1} (\bibinfo {year}
  {2016})}\BibitemShut {NoStop}%
\bibitem [{\citenamefont {Bisi}\ \emph {et~al.}(1955)\citenamefont {Bisi},
  \citenamefont {Germagnoli}, \citenamefont {Zappa},\ and\ \citenamefont
  {Zimmer}}]{Bisi:NC1955}%
  \BibitemOpen
  \bibfield  {author} {\bibinfo {author} {\bibfnamefont {A.}~\bibnamefont
  {Bisi}}, \bibinfo {author} {\bibfnamefont {E.}~\bibnamefont {Germagnoli}},
  \bibinfo {author} {\bibfnamefont {L.}~\bibnamefont {Zappa}}, \ and\ \bibinfo
  {author} {\bibfnamefont {E.}~\bibnamefont {Zimmer}},\ }\href {\doibase
  10.1007/BF02855920} {\bibfield  {journal} {\bibinfo  {journal} {Nuovo
  Cimento}\ }\textbf {\bibinfo {volume} {2}},\ \bibinfo {pages} {290} (\bibinfo
  {year} {1955})}\BibitemShut {NoStop}%
\bibitem [{\citenamefont {Rudstam}(1956)}]{NSR1956RU45}%
  \BibitemOpen
  \bibfield  {author} {\bibinfo {author} {\bibfnamefont {G.}~\bibnamefont
  {Rudstam}},\ }\href@noop {} {\  (\bibinfo {year} {1956})},\ \bibinfo {note}
  {{PhD Thesis, University of Uppsala}}\BibitemShut {NoStop}%
\bibitem [{\citenamefont {Genz}\ \emph {et~al.}(1971)\citenamefont {Genz},
  \citenamefont {Renier}, \citenamefont {Pengra},\ and\ \citenamefont
  {Fink}}]{Genz:1971kv}%
  \BibitemOpen
  \bibfield  {author} {\bibinfo {author} {\bibfnamefont {H.}~\bibnamefont
  {Genz}}, \bibinfo {author} {\bibfnamefont {J.~P.}\ \bibnamefont {Renier}},
  \bibinfo {author} {\bibfnamefont {J.~G.}\ \bibnamefont {Pengra}}, \ and\
  \bibinfo {author} {\bibfnamefont {R.~W.}\ \bibnamefont {Fink}},\ }\href
  {\doibase 10.1103/PhysRevC.3.172} {\bibfield  {journal} {\bibinfo  {journal}
  {Phys. Rev. C}\ }\textbf {\bibinfo {volume} {3}},\ \bibinfo {pages} {172}
  (\bibinfo {year} {1971})}\BibitemShut {NoStop}%
\bibitem [{\citenamefont {Hampel}\ and\ \citenamefont
  {Remsberg}(1985)}]{Hampel:1985zz}%
  \BibitemOpen
  \bibfield  {author} {\bibinfo {author} {\bibfnamefont {W.}~\bibnamefont
  {Hampel}}\ and\ \bibinfo {author} {\bibfnamefont {L.}~\bibnamefont
  {Remsberg}},\ }\href {\doibase 10.1103/PhysRevC.31.666} {\bibfield  {journal}
  {\bibinfo  {journal} {Phys. Rev.}\ }\textbf {\bibinfo {volume} {C31}},\
  \bibinfo {pages} {666} (\bibinfo {year} {1985})}\BibitemShut {NoStop}%
\bibitem [{\citenamefont {D'Agostini}(1994)}]{DAgostini:1993arp}%
  \BibitemOpen
  \bibfield  {author} {\bibinfo {author} {\bibfnamefont {G.}~\bibnamefont
  {D'Agostini}},\ }\href {\doibase 10.1016/0168-9002(94)90719-6} {\bibfield
  {journal} {\bibinfo  {journal} {Nucl. Instrum. Meth.}\ }\textbf {\bibinfo
  {volume} {A346}},\ \bibinfo {pages} {306} (\bibinfo {year}
  {1994})}\BibitemShut {NoStop}%
\end{thebibliography}


\end{document}